\documentclass[12pt,preprint]{aastex}

\usepackage{graphicx}
\usepackage{epstopdf}
\usepackage{color}
\usepackage{enumitem}

\usepackage{natbib}
\def\comm#1 {{\tt (COMMENT: #1) }}

\newcommand{\nhico}{$\rm \frac{{\it N}_{H_2}}{{\it I}_{CO}}$}
\newcommand{\tco}{$T_{\rm CO}$}
\newcommand{\coone}{$\rm ^{12}CO(1-0)$}

\newcommand{\kms}{\>{\rm km}\,{\rm s}^{-1}}

\newcommand{\mum}{\>{\mu {\rm m}}}

\newcommand{\msun}{\>{\rm M_{\odot}}}

\newcommand{\msunpc}{\>{\rm M_{\odot}\,pc^{-2}}}
\newcommand{\dg}{^{\circ}}

\newcommand{\as}{^{\prime\prime}}

\newcommand{\bdm}{\begin{displaymath}}
\newcommand{\edm}{\end{displaymath}}
\newcommand{\beq}{\begin{equation}}
\newcommand{\eeq}{\end{equation}}
\newcommand{\bit}{\begin{itemize}}
\newcommand{\eit}{\end{itemize}}
\newcommand{\ben}{\begin{enumerate}}
\newcommand{\een}{\end{enumerate}}
\newcommand{\bfi}{\begin{figure}[htb]}
\newcommand{\bpfi}{\begin{figure}[p]}

\newcommand{\htwo}{$\rm H_2$}

\newcommand{\ha}{$\rm H\alpha$}
\newcommand{\mhtwo}{$M_{\rm H_2}$}
\newcommand{\shtwo}{$\rm \Sigma_{M_{H_2}}$}

\slugcomment{Accepted version}

\shorttitle{Spiral Arm Cloud and Star Formation}
\shortauthors{Schinnerer et al.}

\begin{document}


\title{The PdBI Arcsecond Whirlpool Survey (PAWS). \\
The Role of Spiral Arms in Cloud and Star Formation}


\author{
Eva Schinnerer\altaffilmark{1},
Sharon E. Meidt\altaffilmark{1},
Dario Colombo\altaffilmark{2},
Rupali Chandar\altaffilmark{3},
Clare L. Dobbs\altaffilmark{4},
Santiago Garc\'{i}a-Burillo\altaffilmark{5},
Annie Hughes\altaffilmark{6},
Adam K. Leroy\altaffilmark{7},
J\'er\^ome Pety\altaffilmark{8,9},
Miguel Querejeta\altaffilmark{1},
Carsten Kramer\altaffilmark{10},
and
Karl F. Schuster\altaffilmark{9}
}

\altaffiltext{1}{MPI for Astronomy, K\"onigstuhl 17, 69117 Heidelberg, Germany}
\altaffiltext{2}{MPI for Radioastronomy, Auf dem HŸgel,  Bonn, Germany}
\altaffiltext{3}{Department of Physics and Astronomy,The University of Toledo, RO 106, Toledo, OH 43606, USA}
\altaffiltext{4}{School of Physics and Astronomy, University of Exeter, Stocker Road, Exeter EX4 4QL, UK}
\altaffiltext{5}{Observatorio Astron\'{o}mico Nacional - OAN, Observatorio de Madrid Alfonso XII, 3, 28014 - Madrid, Spain}
\altaffiltext{6}{IRAP, 9, avenue du Colonel Roche, BP 44346 - 31028 Toulouse cedex 4, France}
\altaffiltext{7}{Department of Astronomy, The Ohio State University, 140 W. 18th Ave., Columbus, OH 43210, USA} 
\altaffiltext{8}{Institut de Radioastronomie Millim\'etrique, 300 Rue de la Piscine, F-38406 Saint Martin d'H\`eres, France}
\altaffiltext{9}{Observatoire de Paris, 61 Avenue de l'Observatoire, F-75014 Paris, France.}
\altaffiltext{10}{Instituto Radioastronom\'{i}a Milim\'{e}trica, Av. Divina Pastora 7, Nucleo Central, 18012 Granada, Spain}


\begin{abstract}
The process that leads to the formation of the bright star forming sites observed
along prominent spiral arms remains elusive. 
We present results of a multi-wavelength study of a spiral arm 
segment in the nearby grand-design spiral galaxy M\,51 that belongs to a spiral density
wave and exhibits nine gas spurs. The combined observations of the
(ionized, atomic, molecular, dusty) interstellar medium (ISM) with star formation
tracers (HII regions, young $\rm <10\,Myr$ stellar clusters) suggest
(1) no variation in giant molecular cloud (GMC) properties between arm and gas
spurs,
(2) gas spurs and extinction feathers arising from the same structure with a close spatial relation 
between gas spurs and ongoing/recent star formation
(despite higher gas surface densities in the spiral arm),
(3) no trend in star formation age either along the arm or along a spur, 
(4) evidence for strong star formation feedback in gas spurs, 
(5) tentative evidence for star formation triggered by stellar feedback for one spur, and
(6) GMC associations (GMAs) 
being no special entities but the result of
blending of gas arm/spur cross-sections in lower resolution observations.
We conclude that 
there is no evidence for a coherent star formation onset mechanism 
that can be solely associated to the presence of the spiral density wave. This suggests that
other (more localized) mechanisms are important to delay star formation such that it occurs in spurs.
The evidence of star formation proceeding over 
several million years within individual spurs implies that the mechanism that leads to star formation acts or is 
sustained over a longer time-scale.

\end{abstract}



\section{Introduction}

The role and importance of spiral arms in the star formation process in galaxy disks is a long-standing
issue. 
Early morphological studies already recognized different
structures associated with pronounced spiral arms emanating, often perpendicular, from the arm.
These features are referred to as spurs if caused by
luminous (stellar) overdensities, feathers when they are due to absorption features and pearls consisting of HII regions 
\citep[for a full account of the history and nomenclature, see e.g.\ introduction of][]{lavigne06}. In particular,
spurs seen as enhancements in (blue) optical light have long been recognized as special locations for the 
formation of stars in these galaxies \citep[e.g.][]{elmegreen80}.
\citet{corder08} presented the first detections of molecular gas line emission coincident with such spurs in
a region of the nearby spiral galaxy M\,51. This strongly suggested that these spurs have counterparts in molecular
gas and thus a strong connection to the star formation process in spiral arms. Their analysis is still the only
study focussing on the properties of gas spurs.

The last decade has seen substantial advances in studying the
ISM in galaxies using numerical simulations. Recent research has
demonstrated the importance of spiral shocks in the formation of
GMCs. In spiral shocks, collisions between clouds occur on relatively
short timescales, allowing clouds to readily coalesce into GMCs \citep[e.g.][]{tan00,kim02,dobbs08}. 
Gravitational interactions between clouds enhance this process,
whilst the higher densities in spiral arms also facilitate
gravitational instabilities \citep{shetty06,dobbs08}.
These numerical models now include heating and cooling of the ISM,
self-gravity, and supernovae feedback \citep[e.g.][]{dobbs11}. 
Furthermore the calculations have sufficient resolution to
predict many properties of clouds such as their masses and virial
parameters \citep[e.g.][]{dobbs08,dobbs11},
as well as the time evolution of GMC and star formation \citep{dobbs13}. The results of these simulations, however, need
constraints from actual observations. 

M\,51 a nearby 
\citep[$\rm {\it D}\sim7.6\,Mpc$;][]{ciardullo02}, almost
face-on \citep[$\rm {\it i}\sim22\dg$;][]{colombo14b} disk galaxy with a clear spiral pattern, provides an excellent
test-bed for theoretical models, and to study the link between spiral
structure and star formation. A close relation between large complexes of GMCs and star formation
sites have been noted already by \citet{vogel88}, and also pointed out for clusters of stellar clusters by
\citet{bastian05}. The inner spiral arm pattern in M\,51 is very likely caused by a density wave, which is present
as perturbation to the gravitational potential of the disk \citep[e.g.][]{tully74, elmegreen89, vogel93, meidt08, colombo14b}
and is thus ideal for a detailed study
of the properties of the interstellar medium (ISM) and star formation across a spiral arm. 
M\,51 has been observed
at essentially all wavelengths and is one of the best-studied grand-design spiral galaxies in the nearby universe.
Recently, high spatial resolution observations of the ISM and star formation activity have been assembled and homogenized
for the PAWS (PdBI Arcsecond Whirlpool Survey) database \citep{sch13}. For our analysis we make use of this database.

The paper is organized as follows. First we briefly describe the data used in \S \ref{sec:data}. 
The molecular gas and star formation
properties across a spiral arm segment are determined in \S \ref{sec:arm}, while we relate
these findings to the star formation process in \S \ref{sec:process}. Implications of our findings
for the general picture of star formation in spiral arms are discussed in \S \ref{sec:discussion}.
We summarize our results and conclude in \S \ref{sec:summary}.


\section{Data}
\label{sec:data}

For our analysis we use the CO(1-0) data products from the PdBI Arcsecond Whirlpool
Survey \citep[PAWS,][]{sch13} tracing the bulk molecular gas in conjunction with ancillary
data probing the other ISM phases as well as different stages of star formation activity.

The molecular gas distribution in the central $\sim$ 9\,kpc of M51 were
obtained as part of the PAWS project \citep{sch13}. This IRAM Large
Program observed 60 pointings with the Plateau de Bure interferometer
(PdBI) in all configurations from 2009 August to 2010 March 
and mapped the full galaxy system with
the 30m single dish telescope in 2010 May in the \coone~ line. The resulting data
cube (with short spacing correction) has a resolution of $1.16\as
\times 0.97\as$ (PA 73$\dg$) with a rms of 0.4\,K per 5$\, \kms$ 
wide channel. In addition, data cubes at 3$\as$ and 6$\as$ resolution were obtained
in a similar fashion, however, using a different weighting of the $uv$ data.
A detailed description of the data reduction is presented by \citet{pety13a}. We 
also use the moment maps derived from
this PAWS datacube \citep[for details see][]{pety13a,colombo14b}.

The ancillary data used and their potential modifications (i.e.\ astrometry) are 
described in detail in section \S 2 of \citet{sch13}. In short, we use the 24$\mu$m image from
Spitzer processed with the HiRes algorithm \citep{dumas11}, the IRAC 8$\mu$m imaging 
from SINGS \citep{kennicutt03} processed by the S$^4$G data pipeline \citep{sheth10} and 
corrected for stellar emission using the ICA method of \citet{meidt12}, the HST $I-H$ map \citep{sch13}, 
the HST ACS $H\alpha$ image from the legacy dataset \citep{mutchler05a} with the prescription 
for continuum correction of \citet{gutierrez11}. In addition, the HI robust weighted data from the THINGS
survey \citep{walter08} and the Herschel PACS [CII] map is used \citep[see][]{sch13,parkin13}.

In addition, we use the catalog of 1,507 GMCs identified in the PAWS area 
\citep{colombo14a}, the catalog of 7,215 grouped HII regions, i.e.\ giant HII regions identified via 
a friends-of-friends algorithm  \citep{lee11}, and the catalog of 3,812 stellar clusters identified in and 
characterized by HST $UVBI+H\alpha$ imaging \citep{chandar16}.


\section{Molecular Gas and Star Formation across a Spiral Arm in M51}
\label{sec:arm}

\begin{figure}
\includegraphics[width=120mm,angle=90]{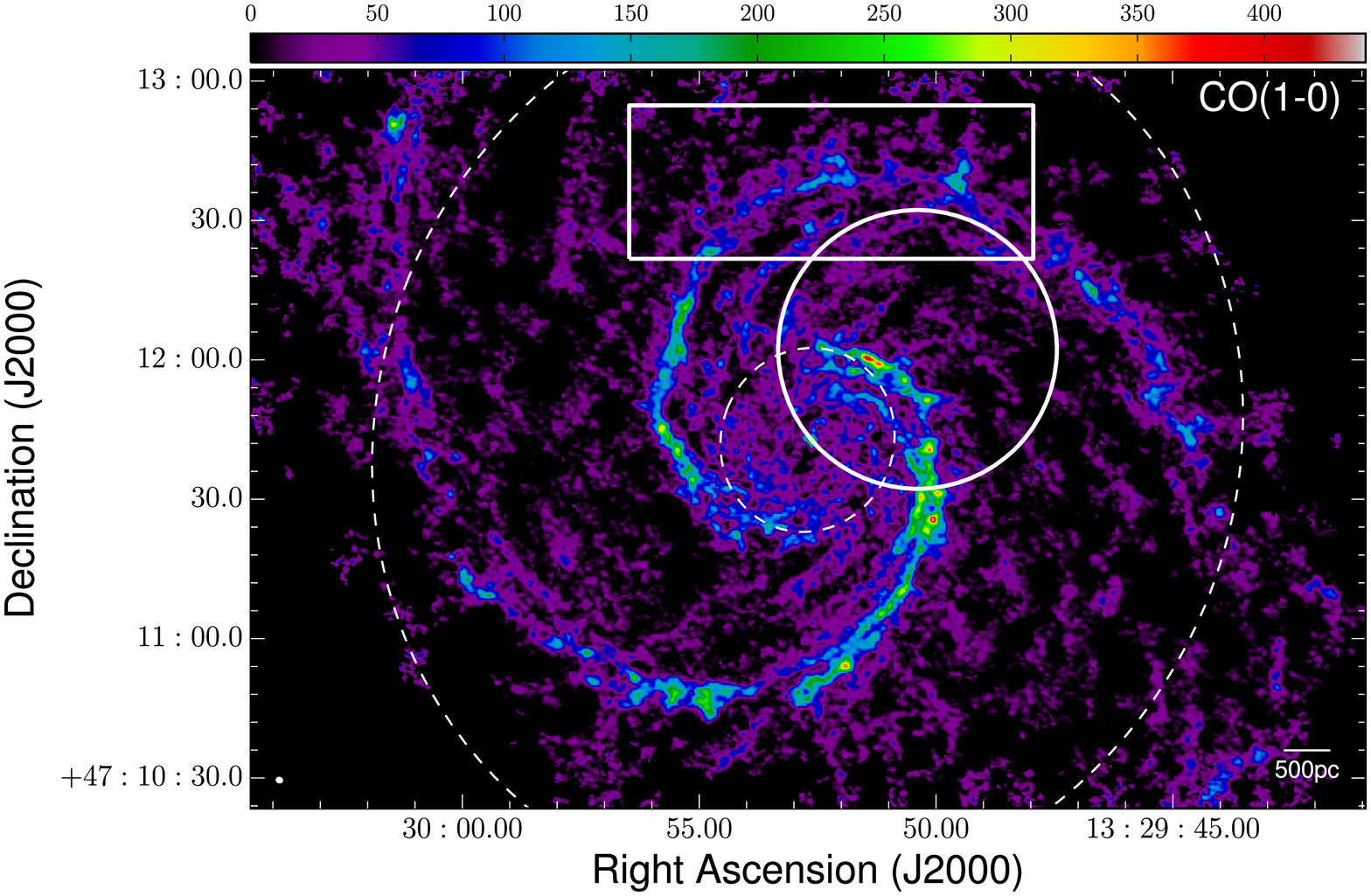}
\caption{Molecular gas distribution as traced by CO(1-0) line emission in the central 9\,kpc of M51a as observed by the PAWS project at $\rm \sim 1\as$
resolution. 
The region studied in detail here is indicated by a rectangular box, the pointing of \citet{corder08} by
a circle. The two dashed circles centered on the nucleus of M51a are at the location of the corotation resonance of the nuclear bar (at $\rm {\it r}\,=\,20\as$) and the {\it m}=2 spiral mode (at $\rm {\it r}\,=\,100\as$) \citep{meidt13a,meidt08}.
\label{fig:co}}
\end{figure}

For the analysis in this paper we focus on a spiral arm region with a bona-fide identification as a spiral 
density wave.\footnote{Note that \citet{corder08} studied a region with zero torque associated with the 
circum-nuclear starburst ring, thus their studied features may have a different formation mechanism than 
the ones discussed here (see Fig.\,\ref{fig:co}).}
Using an enhanced correction for dust emission in the Spitzer/IRAC 3.6$\mu$m image from the S$^4$G pipeline 5 
\citep{querejeta15}, and considering information from beyond the edge of the PAWS FoV, \citet{querejeta16} move 
the co-rotation of the inner spiral pattern to $\rm {\it r}\sim100\,\as$.ÊThis is also consistent with the kinematic decomposition of the 
line-of-sight velocity field traced by the cold ISM as performed by \citet{colombo14b}.Ê That analysis clearly shows that 
the inner spiral pattern has properties consistent with a spiral density wave and reveals the kinematic signature of an 
additional, unique {\it m}=3 mode in the region of  $\rm 20\,\as < {\it r} < 45\,\as$, beyond $\rm {\it r}\sim100\as$ analysis suggest
that the spiral arm is more consistent with a material arm.

As \citet{colombo14b} provide convincing evidence that the superposition of the kinematically 
confirmed {\it m}=3 mode causes an enhancement in the potential of the southern arm and the southern
arm bifurcates twice at $\rm {\it r}\sim45\as$ and $\rm {\it r}\sim65\as$,
we restrict our analysis to $\rm 45\,\as < {\it r} < 75\,\as$ of the northern arm only. This should
also allow for more easy comparison to model predictions.
The location of the area of interest within the PAWS region is presented in Fig.\,\ref{fig:co}.

\subsection{Molecular Gas Properties}
\label{subsec:co}

We utilize the PAWS CO(1-0) moment maps \citep{pety13a,colombo14b} and GMC catalog \citep{colombo14a} 
to study the average properties of the molecular gas 
(see \S \ref{subsec:co_global}) and the GMCs (see \S \ref{subsec:co_gmcs}) in different zones of our selected spiral
arm region. We remind the reader that M\,51's cold ISM is predominantly molecular \citep{schuster07}, especially in the
region under study. Recent
studies of the dust-to-gas ratio in M\,51 \citep{mentuch12} and of the conversion factor between CO(1-0) line intensity and \htwo ~gas mass 
comparing three independent methods (Groves et al. in prep.) show small variations similar to the uncertainties
for SFR tracers at their spatial scales considered ranging from kpc to GMC scale. Note that variations in the conversion 
between CO luminosity and molecular gas mass can be much larger when small regions within individual GMCs are 
considered.

\begin{figure}
\includegraphics[width=120mm,angle=0]{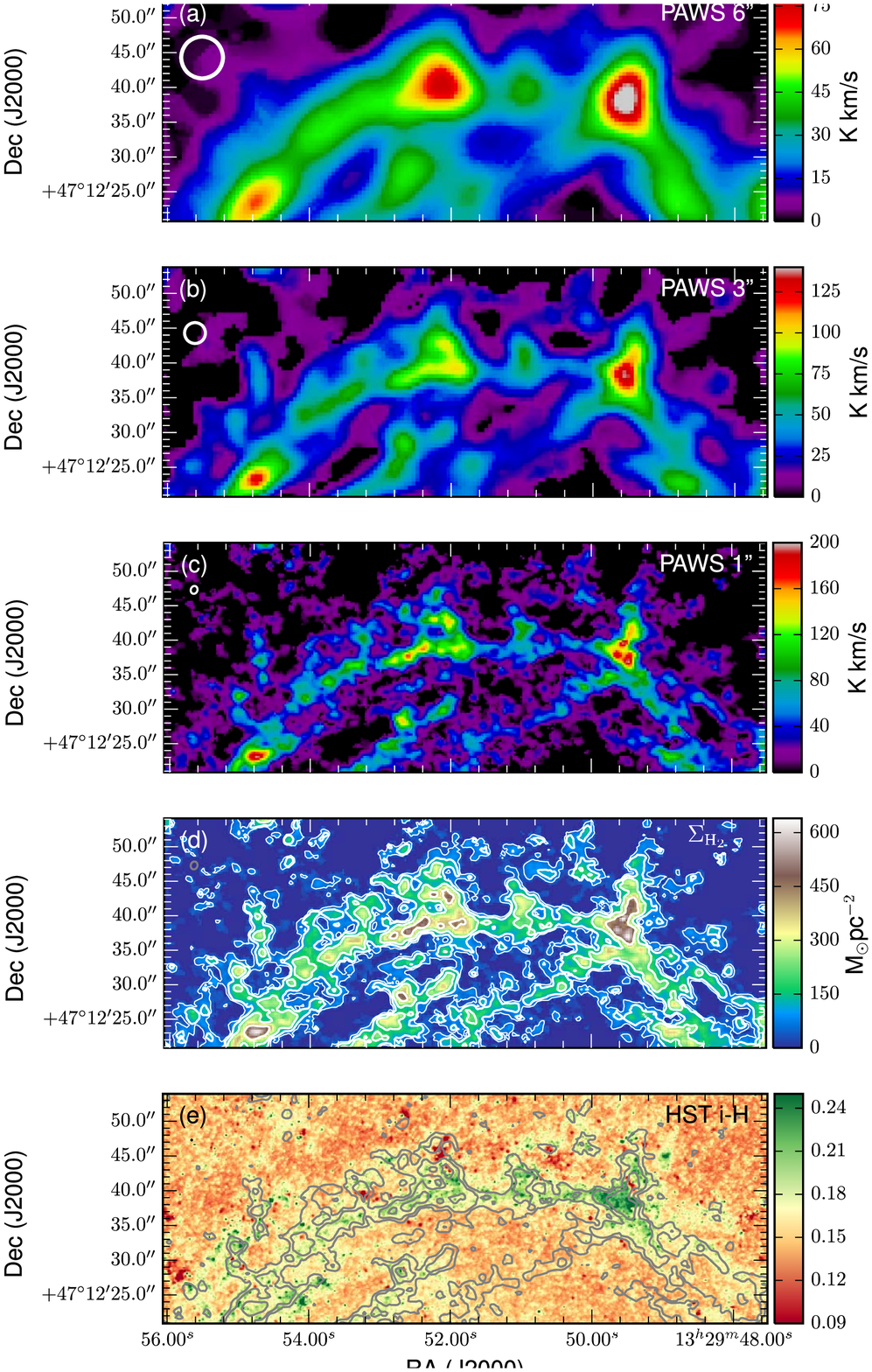}
\caption{Molecular gas distribution as traced by CO(1-0) emission in the northern spiral arm of M51a as seen by PAWS. 
Shown are:
(a) \coone ~intensity distribution at 6'' resolution,
(b) \coone ~intensity distribution at 3'' resolution,
(\,c) \coone ~intensity distribution at $\sim$1'' native resolution with, 
(d) surface density distribution of the molecular gas $\rm \Sigma {\it M}_{H_2}$ with contours of 50 (thin line), 100, 200, 
and 400 $\msunpc$ (thick lines), and
(e) HST $i-H$ color map (green corresponding to the reddest colors) overlaid with gas surface density $\rm \Sigma {\it M}_{H_2}$ contours from panel (d).
\label{fig:spur_co}}
\end{figure}

In order to assess how the distribution/appearance of the molecular gas emission changes as a function of angular 
resolution (Fig.\,\ref{fig:spur_co}), we compare the molecular gas maps as traced by CO(1-0) emission from 6$\as$ to 1$\as$ angular resolution.
The CO(1-0) emission in the spiral arm segment shows five clear peaks with roughly equidistant spacing when mapped at 6$\as$ 
resolution. However, with increasing angular resolution these peaks are resolved into nine gas spurs that 
emanate almost perpendicular from the spiral arm. Comparison to the HST $I-H$ color map reveals an excellent 
coincidence between the molecular gas spurs and the extinction feathers, i.e.\ the extinction features leaving the 
main dust lane along the spiral arm\footnote{For a summary of the naming conventions we refer the reader to the
introduction in \citet{lavigne06}.}. \citet{lavigne06} used HST ACS single band imaging to identify the extinction 
feathers in M\,51. All gas spurs in the spiral arm segment can be matched to their extinction feathers. This finding 
is similar to the result by \citet{corder08} for their spurs emanating from the inner part of the spiral arm that belongs 
to a kinematically different environment.

The good correspondence between the CO emission and the optical extinction as traced by the
HST $I-H$ image (Fig.\,\ref{fig:spur_co}\,d) implies that no obvious spurs have been missed and 
that optical-near-IR color maps are a good predictor for the presence and location of gas spurs.
This, in turn, means that feathers (i.e.\ elongated dust lanes) are caused by the presence of 
dense cold interstellar material and that the gas spurs and the extinction feathers belong to the same structures. 

\subsubsection{Global Molecular Gas Properties in Arm and Spurs}
\label{subsec:co_global}

\begin{figure}
\begin{center}
\includegraphics[width=60mm,angle=-90]{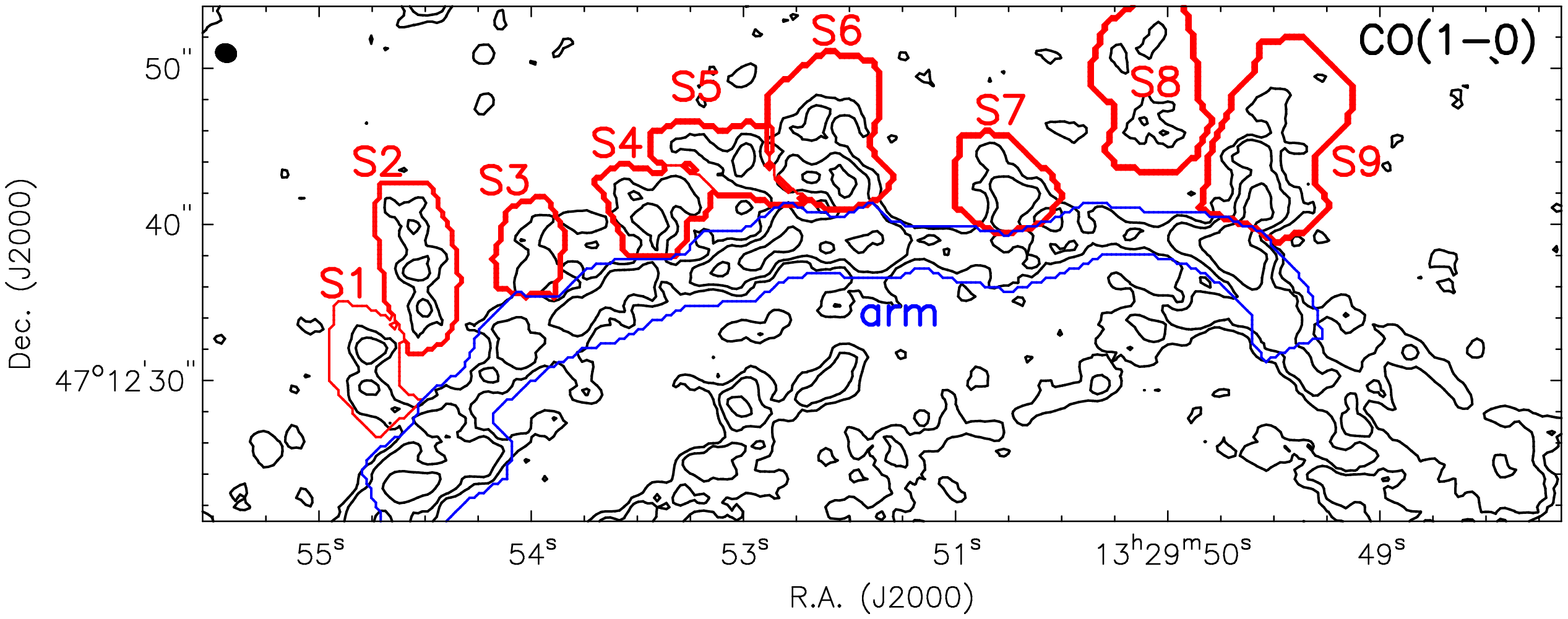}
\end{center}
\caption{Nomenclature of identified gas spurs (red labels) and corresponding segments of the gas spurs (red contour) as well as the corresponding arm segment (blue contour) (see also Tab.\,\ref{tab:co}).
\label{fig:region}}
\end{figure}

\begin{figure}
\includegraphics[width=100mm,angle=0]{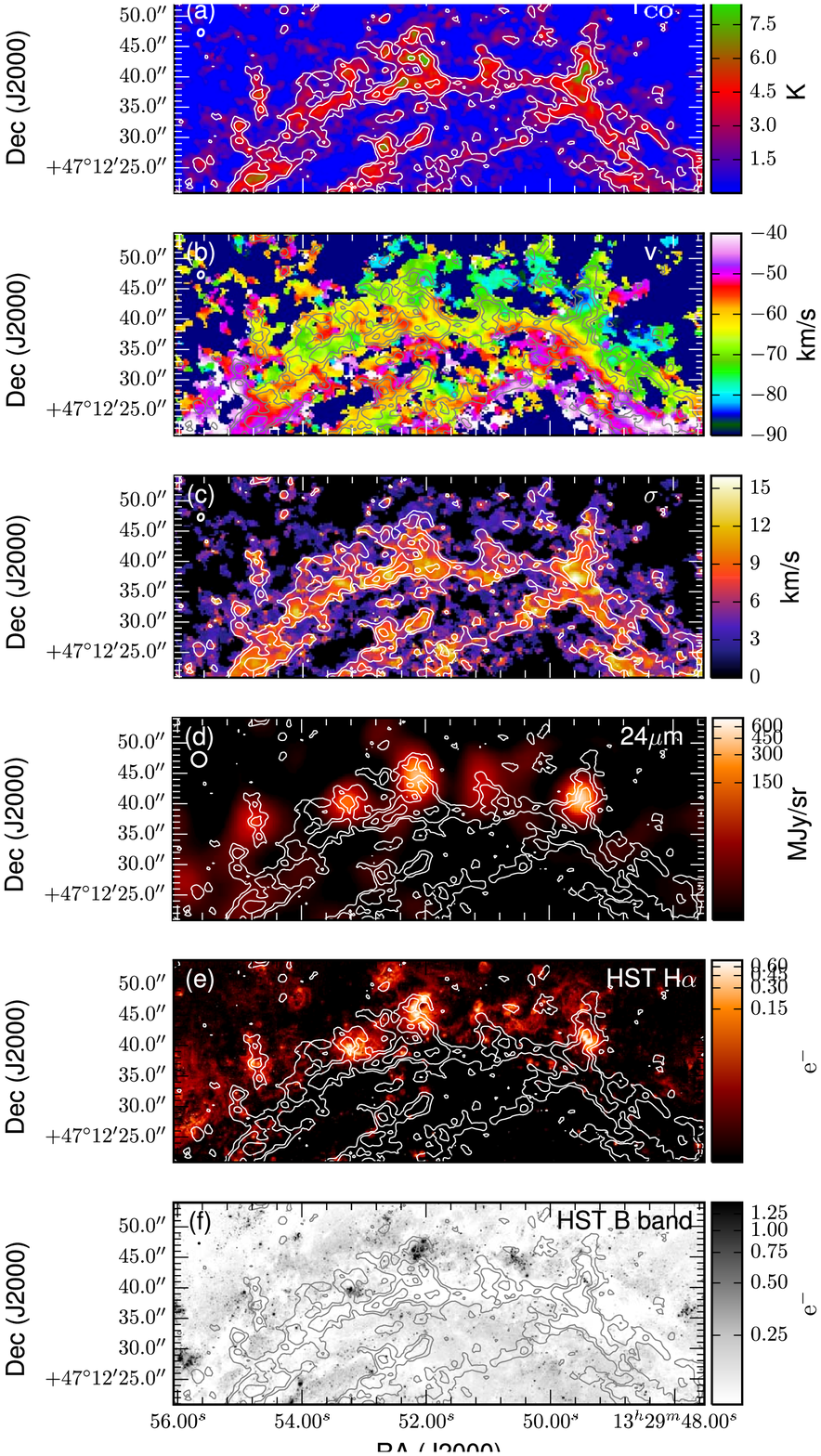}
\caption{Properties of the molecular gas and distribution of star formation tracers in the northern spiral arm of M51a:
(a) \coone ~peak brightness temperature distribution,
(b) \coone ~velocity field colors range from -95 to -45 $\kms$,
(\,c) \coone ~velocity dispersion colors range from 0 to 16 $\kms$,
(d) hot dust emission as traced by MIPS 24$\mu$m continuum,
(e) HII regions traced by HST/ACS $H\alpha$ emission, and
(f) (young) stellar clusters as images in HST/ACS $B$ band continuum.
The contours in all images refer to the molecular gas surface density from Fig.\,\ref{fig:spur_co} (d) (thick lines only).
\label{fig:spur}}
\end{figure}

We identify nine spurs or spur-like features in the northern spiral arm segment 
(see Fig.\,\ref{fig:region} for nomenclature), most of them have counterparts in optical extinction 
feathers as mapped by \citet{lavigne06}. The exceptions are spurs S\,8 which does not connect to the arm
and S\,5 which \citet{lavigne06} associate with S\,6 as a single entity. Seven of the spurs are directly connected to the arm 
at our sensitivity limit while almost all show distinct sub-structures (Fig.\,\ref{fig:spur}\,c). The spacing 
between the bases of the spurs varies between 4$\as$ and 11$\as$ when deprojected 
\citep[using an inclination
of 22$\dg$ and position angle of 173$\dg$; see e.g.][]{colombo14b}. The average spacing is
$\rm \sim 7.5\as$. Their (de-projected) lengths are on average $6.5\as$ with the shortest 
spur S\,4 being about 25\% shorter and the longest spur S\,9 being about 40\% longer.
The (deprojected) spurs have typical widths of 1-2$\as$ (40-80\,pc) in their thin structures and widen up to
several arc-seconds in their thicker parts. The thicker parts are usually offset from the spiral arm.
As the detected line emission is not smooth but exhibits emission peaks, this suggests internal 
structure within the molecular gas spurs. 
Defining regions for each spur and the arm (see red and blue contours in Fig.\,\ref{fig:region}), we measured 
global molecular gas properties such as integrated CO intensity, \htwo ~gas mass 
\mhtwo, the corresponding \htwo ~gas surface density \shtwo ~and the peak brightness 
temperature \tco ~within each spur and the arm (see Tab. \ref{tab:co}). 

In most spurs the CO peak brightness temperature \tco ~is typically as high or even higher in the 
spurs than the adjacent spiral arm segments (Fig.\,\ref{fig:spur}\,a). However, the average (mean
or median)  \tco ~is 
generally about 20-25\% higher in the spiral arm  ($\rm \Delta\,{\it T}_{CO,arm}\,\approx3\,K$) than in the spurs (Tab. \ref{tab:co}). 
The exception is spur S6 which has an equally high mean value. 
Comparison of the mean and median \tco ~shows that the mean is always larger than the median with 
the exception of spur S\,3. While the difference is less than $\sim 5\%$ for the arm and spurs S\,4 and S\,5,
it is around 10\% for the remaining spurs except for spurs S\,3 and S\,9 where the difference is about 20\%.
This suggests that the distribution of \tco ~is skewed towards lower values in the majority of the spurs.
It is interesting to note that the maximum 
\tco ~value in two spurs (S6, S9) is higher than the maximum \tco ~measured in the arm.

Using the integrated CO line emission the picture reverses and the spiral arm becomes 
significantly brighter, also relative to most spurs (e.g.\ Fig.\,\ref{fig:spur}\,a). 
Assuming the standard Galactic conversion factor between CO luminosity and molecular 
hydrogen mass of \nhico $\rm = 2\times10^{20}\,cm^{-2}K^{-1}km^{-1}s$ we calculated the distribution 
of the \htwo ~surface density \shtwo. In the spurs we find an average \htwo ~gas surface 
density of \shtwo $\approx 100\,\msunpc$ (twice the value found by \citet{corder08} for their 
spurs) reaching maximum values of up to 
$400-500\,\msunpc$ in the more prominent spurs (S\,6 and S\,9) (see Tab. \ref{tab:co}).
However, we do not see a good correlation between the \htwo ~surface density in the spurs and 
in the immediate adjacent spiral arm segments (Fig.\,\ref{fig:spur_co}\,d) as \shtwo ~has both lower and higher
values in the spurs than in the corresponding arm segment.

Taken together these results imply that although the gas surface density \shtwo ~is higher in the arm, 
the gas in the spurs is on average brighter based on the lower contrast 
between arm and spurs in the peak brightness temperature. This brightness increase could be caused
by a higher filling factor, a higher gas volume density or a higher gas kinematic temperature.
From this finding (similar \tco, but different \shtwo) immediately follows
that the velocity dispersion $\rm \sigma$ in the spurs is on average lower than $\rm \sigma$ in 
the spiral arms (see Fig.\,\ref{fig:spur}\,d and Tab. \ref{tab:co}). 
The value for the spurs is on average a third lower than the arm value 
of $\rm \sigma=8.1\,km/s$. The typical gas mass within a 
spur is \mhtwo\,$\rm \approx\,4\times\,10^{6}\,\msun$ slightly higher than the amount of gas 
present in the spurs of \citet{corder08} (when correcting for the different distance used). The 
spurs S\,6 and S\,9 contain about three times as much gas and cover both some of the largest area.

The spurs form a distinct kinematic environment as the gas associated with the spurs shows 
strong deviations from the regular velocity field with velocity gradients (of roughly $\rm 5\,km/s$
per arc second) along the minor
axis of the spurs, e.g.\ S\,1, S\,3, S\,4, S\,6, and S\,9 (see Fig.\,\ref{fig:spur}\,b), though the direction
of the gradient is not always the same. These strong 
streaming motions are not similar to the ones seen in the spiral arm itself where the velocity 
gradient is generally largest across the arm width, so almost perpendicular to the gradient seen 
across the spurs.

\subsubsection{Giant Molecular Cloud (GMC) Properties in Arm and Spurs}
\label{subsec:co_gmcs}

In order to study the properties of GMCs in our defined spur and arm regions (see red and blue 
contours in Fig.\,\ref{fig:region}) we utilize the PAWS GMC catalog \citep{colombo14a} of GMCs 
identified via the CPROPS software \citep{rosolowsky06}. In Fig.\,\ref{fig:spur_obj}\,a we highlight all 
GMCs associated with a spur (arm) as red (blue) circles. GMCs are mostly associated with the 
arm and the spurs; larger GMCs preferentially coincide with gas over-densities. A summary of 
the average properties of all GMCs found in either the arm or spur regions are listed in 
Tab. \ref{tab:gmcs_sum}. (The individual properties of all identified GMCs are provided in 
Tab. \ref{tab:gmcs} and \ref{tab:gmcs_arm}.)

The number of GMCs identified in the gas spurs ranges from 1 to 6, with an average of 2.8 GMCs 
per spur. Spur GMCs have sizes of $\rm {\it r}\sim50\,pc$ similar to the average arm GMC when the 
two large GMCs of spurs S\,4 and S\,8 are excluded. The typical line-width of spur GMCs is slightly 
lower than for arm GMCs consistent with the observed lower velocity dispersion in the spurs. 
The typical molecular gas mass of spur GMCs is about $\rm 3\times10^6\,\msun$ and 
therefore the typical spur GMC is about 25\% more massive than the average arm GMC. The amount of molecular gas 
in GMCs per defined region (i.e.\ spur/arm) is still lower in the spurs than the arms, the difference is 
now only a factor of 1.5$\times$ compared to two times for the gas surface density. 
Interestingly,
the average fraction of gas mass in GMCs versus total gas mass per region is close to unity ($\rm \sim 95\%$)
for the spurs\footnote{The reason that for some spurs more gas is found in GMCs compared to the spur area 
is due to the fact that the GMCs can extend beyond the spur area identified.}, while only about $\rm \sim75\%$ of
the molecular gas mass in the arm segment can be allocated to GMCs (see Tab. \ref{tab:gmcs_sum}).

In summary, it appears that GMCs in the arm and spurs have fairly similar properties with a slight
preference for spur GMCs to be more massive. At the same time there is an indication that most molecular
gas in the spurs is in GMCs while a significant fraction of the gas in the arm might be distributed in
less coherent (potentially more unbound) structures.

\subsection{Star Formation Properties}
\label{subsec:sf}

The presence of massive star formation relative to the molecular gas distribution is investigated
using three star formation rate tracers, namely hot dust, ionized gas and blue optical light (see \S
\ref{subsec:sf_tracers}). A quantitative analysis of star formation in the different zones of the
selected spiral arm region is made utilizing catalogs of (giant) HII regions \citep{lee11} and young
($\rm \le 10\,Myr$) stellar clusters \citep{chandar16} identified in HST images (see \S \ref{subsec:sf_cats}).

\subsubsection{Massive Star Formation in Arm and Spurs}
\label{subsec:sf_tracers}

The location of ongoing and recent massive star formation relative to the molecular gas arm and its
spurs is shown in Fig.\,\ref{fig:spur}\,d-f. All three star formation tracers 
($\rm 24\,\mum$ emission from hot dust heated by massive stars, \ha ~emission from HII regions, 
and blue clusters of young stars in a HST $B$ band image) show that (massive) star formation is almost exclusively 
associated with the gas spurs and that no prominent star formation is taking place in the arm itself. 

All of the five prominent $\rm 24\,\mum$ emission peaks that can be identified in our selected region 
(Fig.\,\ref{fig:spur}\,d) coincide with gas spurs (S\,2, S\,4, S\,6, S\,7, and S\,9) with the two brightest ones 
coinciding with spurs S\,6 and S\,9 that contain most of the molecular gas as traced by CO emission. The enhanced $\rm 24\,\mum$ 
emission associated with the arm itself is very likely not due to embedded star formation but rather due 
to the higher gas density as discussed by \citet{sch13}. Given the sensitivity of the  data 
there is no evidence for highly embedded star formation occurring in the arm.
When using the full PAWS area and focussing on the brightest $\rm 24\,\mum$ peaks 
($\rm 20\,MJy/sr\,<\,{\it S}_{24\mum}\,<\,200\,MJy/sr$), we find that indeed most peaks coincide with gas spurs.  
Excluding the central ring-like area, the numbers are 9/11 peaks in the southern arm and 7/9 peaks 
in the northern arm. 

The \ha ~emission (Fig.\,\ref{fig:spur}\,e) is abundant north of the molecular gas arm and mainly arises 
from large (up to 5'', i.e.\ 185\,pc, diameter) shell-like structures, consistent with HII regions of ages of 
5-10\,Myrs \citep{whitmore11}. All bright and large \ha ~emitting regions coincide with spurs 
(S\,2, S\,4, S\,6, and S\,9) while fainter emission is associated with three more spurs (S\,1, S\,7, 
and possibly S\,5). Interestingly, the location of the large \ha ~regions within the spurs varies from 
being close to the base of the spur (e.g., S\,3) to its tip (e.g., S\,6). Most of the gas arm itself is free 
of  prominent \ha ~emission, except for the segments adjacent to spur S\,5 and S\,1.
Over the entire PAWS area, basically all large HII region complexes are associated with gas 
spurs. We find, again excluding the central ring area, 11/12 in the northern arm and 7/7 in the southern arm.

The HST $B$ band image shows the distribution of (young) stellar clusters relative to the molecular 
gas arm and its spurs (Fig.\,\ref{fig:spur}\,f). Three prominent stellar cluster complexes are evident. 
Two of them coincide with spurs (S\,4,  S\,6) while one lies off the arm roughly between spur 
S\,7 and S\,8. No prominent clusters are obvious within the gas arm while most
clusters are seen associated with spurs.

In summary, all three star formation rate tracers sensitive to the sequence from embedded star 
formation to stellar clusters devoid of their birth clouds are almost exclusively associated with gas spurs.
This finding is not necessarily surprising as already \citet{elmegreen80} noted that feathers, i.e.\ dust 
lanes corresponding to the gas spurs, are interspersed by pearls, i.e.\ large HII regions. This 
strongly suggests that there might be a causal link between gas spurs and massive star formation. 
We find no evidence for prominent ongoing star formation in the molecular gas arm itself.

\subsubsection{HII Regions and Young ($\rm \le 10\,Myrs$) Stellar Clusters in Arm and Spurs}
\label{subsec:sf_cats}

We utilize the grouped HII region catalog of \citet{lee11} that is based on the HST Heritage Program 
imaging of M\,51. The authors have corrected the \ha ~flux for [NII] contamination, Galactic foreground 
and an intrinsic mean attenuation of $\rm {\it A}_V \approx 3.1$ \citep{scoville01}.\footnote{We corrected 
the catalog entries to our adopted distance.} In addition, they identified giant and super-giant HII 
regions via a friends-of-friends algorithm. We associate an HII region (of 
$\rm log({\it L}_{H\alpha}(erg\,s^{-1}) \ge 37.0$) with a spur or the arm if its 
center falls within our defined regions. The average HII region properties are summarized in 
Tab. \ref{tab:hii_sum} while Tab. \ref{tab:hii} lists all the associated HII regions from the catalog 
of \citet{lee11} corrected to our adopted distance and astrometry \citep[for details on the astrometry, 
see][]{sch13}.

\begin{figure}[ht]
\begin{center}
\resizebox{.6\hsize}{!}{\includegraphics[angle=-90]{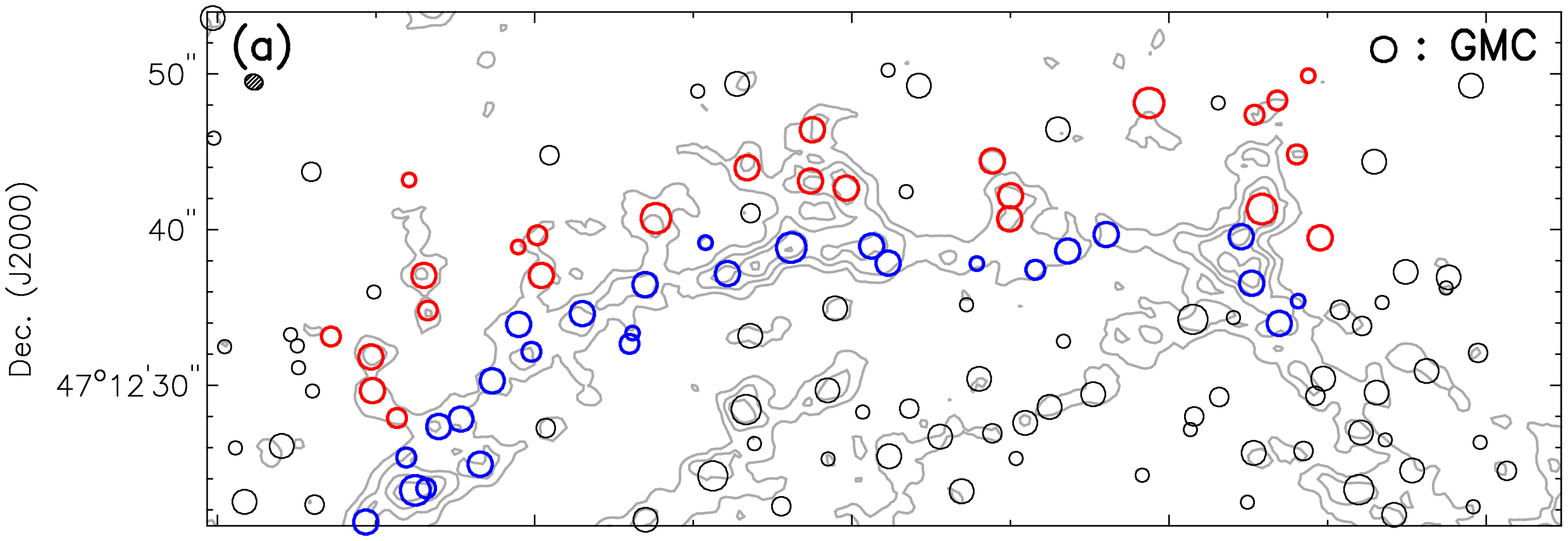}}\\
\resizebox{.6\hsize}{!}{\includegraphics[angle=-90]{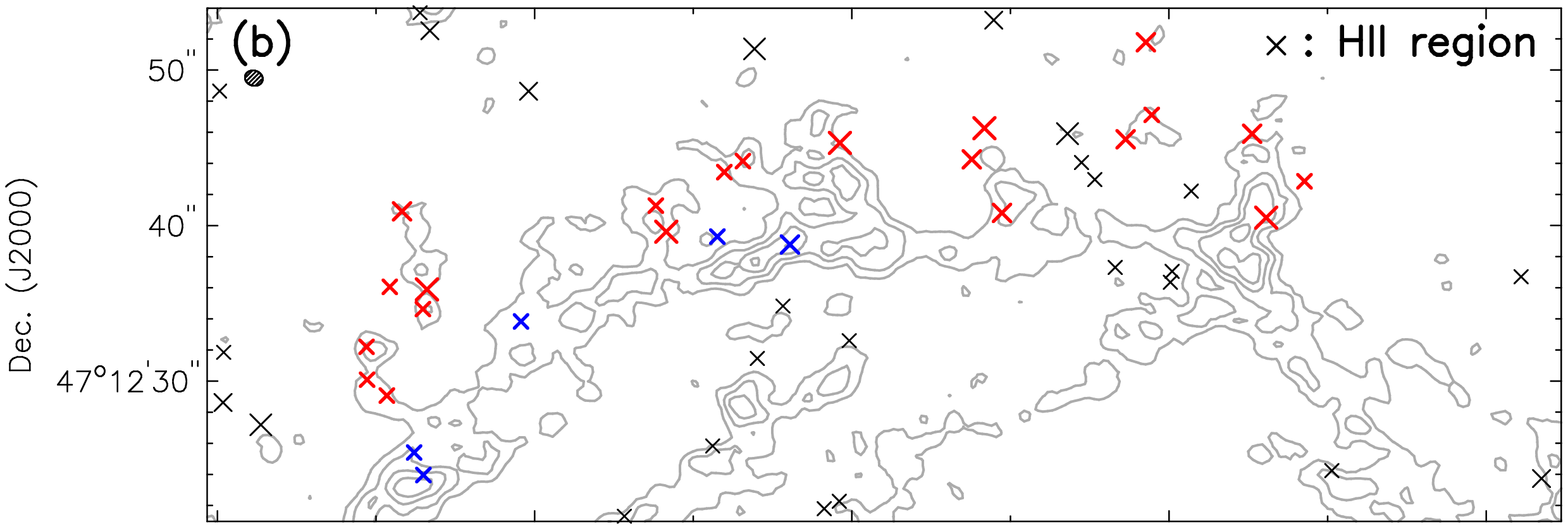}}\\
\resizebox{.6\hsize}{!}{\includegraphics[angle=-90]{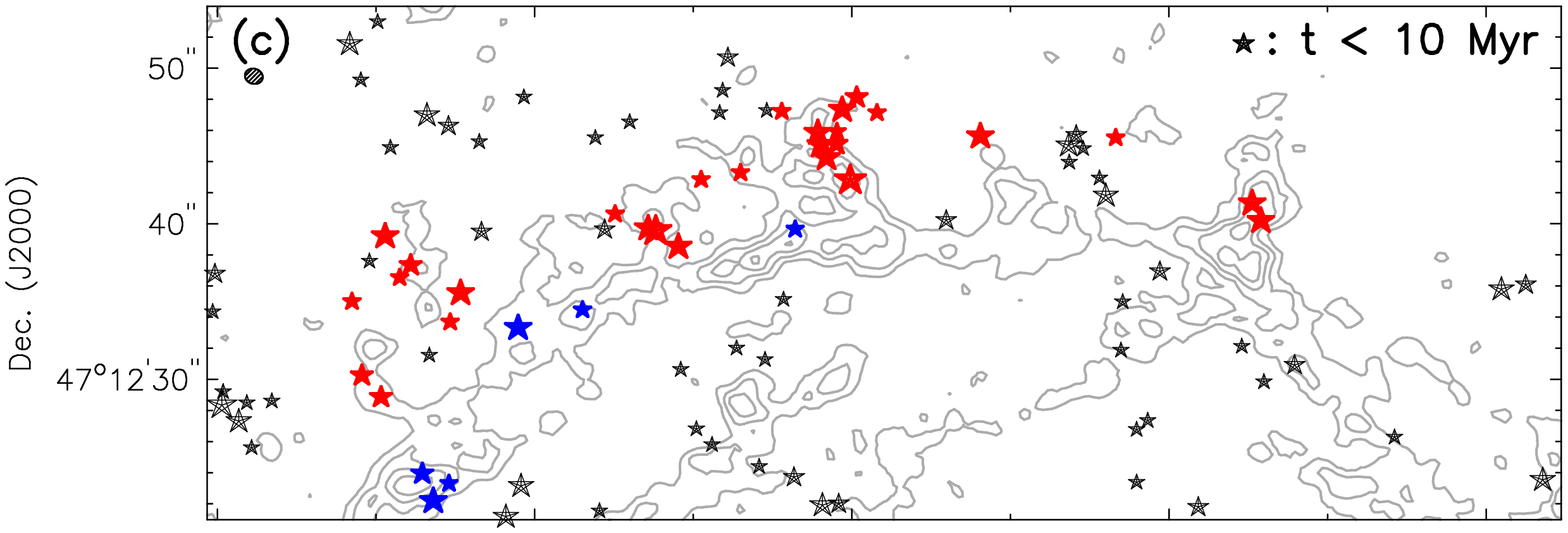}}\\
\resizebox{.6\hsize}{!}{\includegraphics[angle=-90]{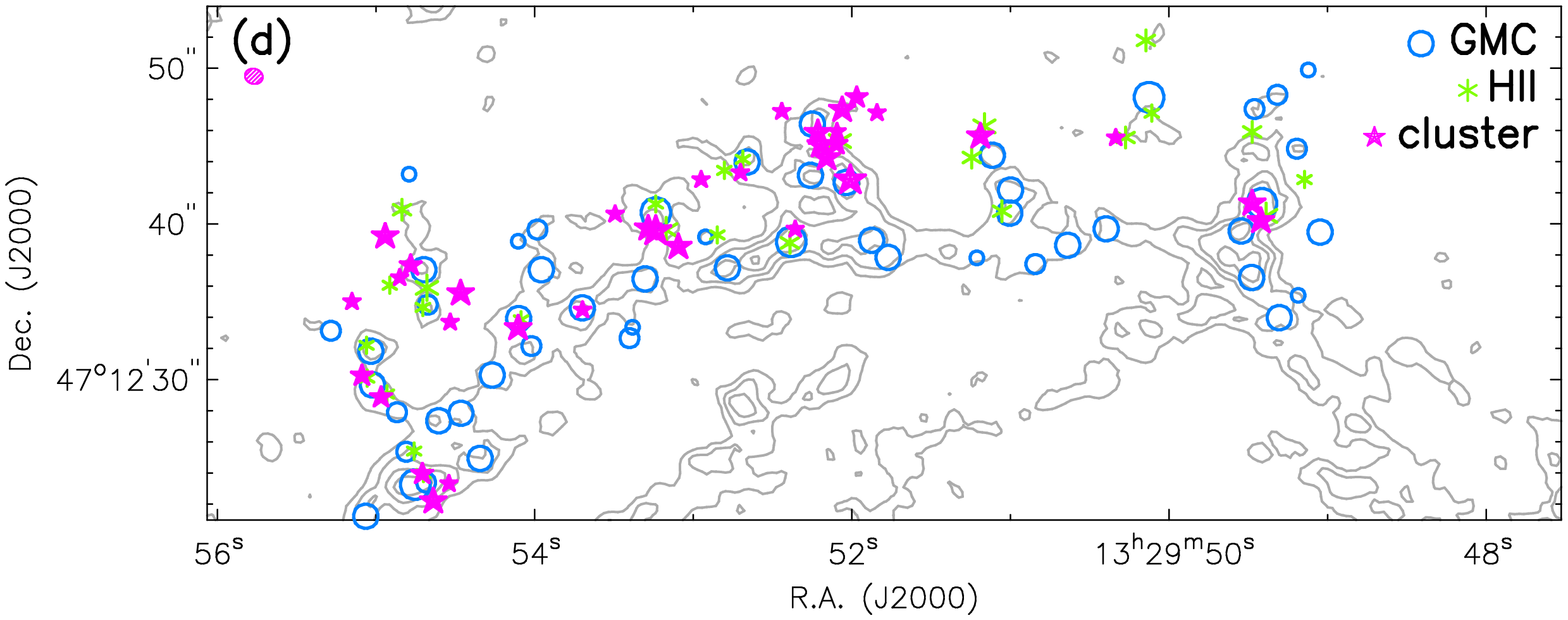}}
\end{center}
\caption{Location of identified GMCs \citep[from][]{colombo14a}, HII regions \citep[from][]{lee11} and young stellar clusters \citep[from][]{chandar16} present in the region under study. Objects residing in our gas spiral arm ({\it blue}) and gas spurs ({\it red}) are color-coded:
(a) GMCs from the catalog of \citet{colombo14a}, the size of the symbols corresponds to their respective mass 
(from small to large: 
$\rm {\it M}_{GMC} < 5\times10^5\,\msun, 
5\times10^5\,\msun \ge {\it M}_{GMC} < 1\times10^6\,\msun, 
1\times10^6\,\msun \ge {\it M}_{GMC} < 5\times10^6\,\msun,
{\it M}_{GMC} > 5\times10^6\,\msun$),
(b) HII regions from the catalog of \citet{lee11}, the size of the symbols corresponds to their respective \ha ~luminosity
(from small to large:
$\rm log({\it L}_{H\alpha}) < 37.5,
37.5 \le log({\it L}_{H\alpha}) < 39.0,
log({\it L}_{H\alpha}) \ge 39.0$), and 
(c) stellar clusters younger than 10\,Myr from the catalog of \citet{chandar16}, the size of the symbols 
corresponds to their respective stellar mass (from small to large:
$\rm {\it M}_{\star} < 5\times10^3\,\msun,
5\times10^3\,\msun \le {\it M}_{\star} < 1\times10^4\,\msun,
1\times10^4\,\msun \le {\it M}_{\star} < 5\times10^4\,\msun,
{\it M}_{\star} > 5\times10^4\,\msun$).
Three catalogs are combined in the last panel (d). 
\label{fig:spur_obj}}
\end{figure}

As expected from the distribution of  the \ha ~line emission most (21/26) of the grouped HII 
regions of \citet{lee11} belong to spurs (see Fig.\,\ref{fig:spur_obj}\,b and Tab. \ref{tab:hii_sum}) despite the fact that 
the spurs and the arm encompass roughly similar areas. On average the HII regions in the 
spurs are twice as large with a radius of $\rm {\it r}\,\approx\,70\,pc$ and over one order of magnitude 
brighter ($\rm {\it L}_{H\alpha} \approx 6\times10^{38}\,erg\,s^{-1}$) than the HII regions in the arm. 
Consequently, the \ha ~luminosity normalized by area in the spurs is about 30 times higher 
than in the arm. Of the four super-giant HII regions with $\rm log({\it L}_{H\alpha}(erg\,s^{-1}))\,>\,39.0$, 
similar to 30\,Doradus in the LMC or NGC\,604 in M\,33, present in the area under investigation, 
three reside in spurs (S\,4, S\,6, S\,9) while the remaining one is located north of spur S\,9. Of 
the nine HII regions with $\rm 37.5\,\le\,log({\it L}_{H\alpha}(erg\,s^{-1}))\,<\,39.0$ present in our defined region, 
eight are located in spurs and one in the arm (next to spur S\,5 in the middle of the spiral arm), while several more 
are found north of the arm and spurs with a typical distance of about 10''.  It is interesting to note that one spur 
(S\,3) does not contain any HII regions.

About 3,500 stellar clusters with a 90\% completeness level down to  
$m_V \approx 23$ have been identified in a summed HST $BVI$ image by applying certain selection criteria on 
morphology \citep{chandar11,chandar16}. The ages and masses are estimated using $U, B, V, I$ and 
(non-continuum-subtracted) $H\alpha$ aperture photometry together with population synthesis 
models \citep[for a more detailed discussion of the young stellar clusters, see][]{calzetti10}. 
We select all stellar clusters with derived ages of $\rm {\it t}\,\le\,10\,Myr$ from the catalog that are 
located in our area of interest. Again we associated these young stellar clusters with spurs and 
the arm if their position falls within our defined regions. The average properties of these 
clusters are summarized in Tab. \ref{tab:cluster_sum} while the individual cluster properties are 
listed in Tab. \ref{tab:clusters}. All values have been corrected to our adopted distance.

We associate a total of 28 young stellar clusters with the spurs (see Fig.\,\ref{fig:spur_obj}\,c). 
One third is found in spur S\,6 alone, one sixth in spur S\,2 and S\,4 each, none in spur S\,3, the 
remaining third is distributed in the other five spurs. A group of six stellar clusters is located 
between the two spurs S\,7 and S\,8 about 4'' north of the gas arm. The six stellar clusters found in the arm are co-spatial 
with the HII regions and next to spurs S\,1, S\,3 and S\,5. There is a preference for the more 
massive stellar clusters to be associated with the spurs. The stellar clusters within the spurs are 
not uniformly distributed but tend to cluster together. This behavior has already been noted by 
\citet{bastian05} who identified complexes of stellar clusters within M\,51. The stellar clusters 
associated with spurs S\,6 and S\,8 correspond to their complexes C2 and D2, which are both 
consistent with homogeneous young stellar populations. Based on the high derived star 
formation rate surface density \citet{bastian05} classify C2 as a localized starburst. The 
difference in the spatial distribution of young stellar clusters between spurs and arm is also 
evident in the amount of (young) stellar mass per area, where the density is almost an order of 
magnitude higher for the spurs compared to the arm.

While the average age of the young stellar clusters in the spurs is about 50\% lower than in the 
arm, this is still within the uncertainty of a factor of 2 \citep{calzetti10}. However, it is 
interesting that the fraction of the youngest clusters with ages of $\rm {\it t} < 3\,Myr$ in the spurs 
(30\% or 9 out of 28) is about twice as high as for the arm (17\% or 1 out of 6). The young stellar 
clusters in the spurs are on average twice as massive compared to those in the arm, which is 
significant given the uncertainty on the stellar mass of $\approx 60\%$ \citep{calzetti10}. 
Subsequently, the surface density of stellar mass contained in young stellar clusters is over 
four times higher in the spurs compared to the arm. As these systems should all have 
basically evacuated their surrounding birth material, i.e.\ the dust and gas that could 
potentially attenuate the stellar light, the difference between spurs and arm cannot be 
explained by significantly higher extinction affecting the arm stellar clusters. Unless one
invokes a significantly different scale height for gas and dust in the arm compared to the
spurs.

Similarly to the distribution of the SFR tracers (see Section \ref{subsec:sf_tracers}) most of the
HII regions and young stellar clusters are associated with the gas spurs. Interestingly, the four 
spurs that harbor the most luminous HII regions contain on average more massive young 
stellar clusters. The arm star formation sites are all close to spurs that only contain less 
luminous HII regions and less massive young stellar clusters (S\,1, S\,5) or even none at all 
(S\,3).


\section{The Star Formation Process in Gas Spurs}
\label{sec:process}

\subsection{Relation of Spurs and Giant Molecular Cloud Associations (GMAs)}
\label{subsec:co_gmas}

The concept of Giant Molecular Associations (GMAs) was first introduced by \citet{vogel88} based on low-resolution
interferometric observations of the molecular gas in the spiral arms of M\,51. These authors suggested that GMAs 
are formed out of GMCs that are already primed for star formation and that their formation is promoted by spiral 
density waves. A detailed follow-up study by \citet{rand90} using similar CO data for most of the molecular gas 
disk in M\,51 found that 20 out of their 26 identified GMAs with molecular gas masses of 
$\rm (0.4 - 2.5)\times10^7\,\msun$ \footnote{Values corrected to our assumed distance of 7.6 Mpc and a Galactic 
conversion factor of $\rm 2\times10^{10}\,cm^{-2}K^{-1}km^{-1}s$} reside in the spiral arms and appear at their 
resolution of $\rm 10\as \times 7\as$  to be virially bound. In the literature it has become common to refer to 
coherent molecular gas structures above $\rm 10^7\,\msun$ as GMAs. For
example, using their high angular 4$\as$ resolution CO imaging of M\,51, \citet{koda09} identify several GMAs 
with molecular gas masses of $\rm {\it M}_{mol} > 10^7\,M_{\odot}$ associated with the spiral arm segment under 
study.

In order to assess the interpretation and identification of GMAs in M\,51 we use our findings from \S \ref{subsec:co} where
we investigated how the distribution/appearance 
of the molecular gas emission changes as a function of angular resolution (see Fig.\,\ref{fig:spur_co}). The two most
prominent emission peaks at $\rm 6\as$ correspond to GMA A1 and A2 from \citet{rand90} and can be associated to
our spurs S\,5 plus S\,6 and S\,9, respectively. Inspecting the location of the twelve arm-GMAs cataloged by \citet{rand90} that fall
within the full PAWS FoV, we find that 9 can clearly be associated with gas spurs (GMAs A1, A2, A3, A6, A8, A9, A10, A12, A13). 
One GMA coincides with a feather \citep[][though no
clear gas spur is seen in this location that is close to the edge of the PAWS map]{lavigne06}. For GMA A4 it is hard to associate
it with a clear spur/feather signature as it is located in the arm region of suppressed star formation \citep[e.g.][]{sch13,meidt13a}, while
GMA A11 falls into the inter-arm region. The mass of all GMCs associated with spurs S\,5 and S\,6 amounts to 
$\rm 1.6\times10^7\,\msun$, i.e.\ close to the mass of GMA A1, the GMCs of spur S\,9 have a total of $\rm 8.7\times10^6\,\msun$
that corresponds to $\sim 40\%$ of the molecular gas mass of GMA A2. As emission
from the neighboring arm segment must have contributed to the total flux determined for the GMAs 
in the lower resolution imaging, this suggests
that the GMAs are actually the blurred combination of gas in the spurs plus their neighboring arm segments, i.e.\ due to
the low resolution these 'cross sections' appear bright.

Based on the analysis above, we conclude that the GMAs identified by \citet{rand90} in M\,51 are most likely an artifact of 
low-resolution observations where 
spatially separate emission from spurs and their neighboring arm are blended together. GMAs are therefore probably not single 
or special entities of multiple GMCs. This finding is not necessarily in disagreement with results of the high-angular resolution study of a 
segment of M51's southern spiral arm \citep{egusa11} where the authors found a higher density of GMCs at the location of 
more massive GMAs identified by \citet{koda09} and interpreted this as evidence that GMAs are smooth structures
that break up into collections of GMCs. As that high resolution study missed 90\% of the emission, it is difficult to link
it directly to the fainter spurs that were identified in HST imaging \citep{lavigne06}.

\subsection{(Relative) Age of Star Formation Activity}
\label{subsec:gas_sf}

We use the spatial distribution of the products of star formation such as
hot dust/PAH (via its $\rm 24\,\mu$m and non-stellar $\rm 8\,\mu$m emission), ionized and
atomic hydrogen (\ha ~and HI emission) as well as ionized carbon ([CII]$\rm \lambda 158\mu m$ emission) in
conjunction with HII regions (including 
the morphology of the \ha ~emission) and young stellar clusters (see HST $B$ band) relative to the 
distribution of the molecular gas (traced via CO) to infer the relative age of the star formation activity associated 
with the spurs and arm as well as the typical location of star formation sites relative to the arm. This 
information 
allows us to search for an age trend which would be expected if, for example, all star formation would be started at 
the same location (e.g.\ inside the arm). We note that the prominence (as traced by their brightness) 
of the star formation sites varies considerably across the spurs and could introduce some bias in our 
age assignment. 

\subsubsection{Along the spiral arm}

As spur S\,3 contains abundant molecular gas, but it shows no sign for any associated star formation nor any evidence of 
star formation impact, we consider it a potential site for future star formation, i.e.\ the relative 'youngest' among our 9 spurs. 

Both spurs S\,7 and 9 have prominent 
$\rm 24\,\mu$m and \ha ~emission with significantly fewer $B$ band clusters, suggesting that massive star formation has 
only recently started in these spurs. Spur S\,7 might be in a slightly earlier star formation phase than spur 
S9 given the faintness of the clusters in the $B$ band. This interpretation is consistent with the distribution of the
ISM dissociation products (HI and [CII] line
emission): an HI (but no [CII]) peak is associated with S\,7, while both emission lines
are mainly found downstream of the major star forming site in S\,9 (i.e.\ peaks in H$\alpha$ and hot dust/PAH emission).
The most prominent, ongoing, massive star 
formation sites are associated with spurs S\,4 and 6 where all three star formation indicators 
are co-spatial with the CO emission. The difference in the [CII] over HI ratio between the two spurs
could indicate that star formation might have been proceeding for a longer time in spur S\,6.
It is also interesting 
to note that the spatial coincidence of $\rm 24\,\mum$, \ha ~emission and young stellar clusters implies 
that massive star formation has occurred within a spur over several Myr, e.g.\ spurs S\,2, S\,4, S\,6, S\.9.

Similar to the previous spurs, spur S\,2 has also all three star formation tracers 
associated with it, however, they are clearly spatially separated. The young clusters tend to be located on the edge of 
the gas spur, suggesting that star formation has been proceeding for a while already inside this spur. 
The situation in spur S\,1 is similar, though the $\rm 24\,\mum$ emission is significantly reduced 
implying that this star formation site is slightly older. The lack of associated $\rm 24\,\mu m$ emission 
together with significant \ha ~emission and young stellar cluster suggests that the formation of massive 
stars has just ceased in spur S\,5.

The over-density of stellar clusters that is located between the arm 
and spur S\,8 is the relative 'oldest' star forming site as no $\rm 24\,\mum$ emission is present and the 
\ha ~emission is very faint and diffuse, suggesting that the HII region has already dissolved. 
The presence of significant [CII] emission suggests that there is still a significant amount of ionizing photons
from massive stars and/or that the recombination time for $\rm C^+$ is much longer than for $\rm H^+$ to HI
to \htwo. Similarly,
the distribution of the molecular gas as seen in the CO emission is more dispersed, suggesting that
the previous massive star formation events have had a severe impact on the morphology and 
prominence of the spur. It also shows
that a spur is not necessarily continuously fed by new material from the arm.

\begin{figure}[ht]
\begin{center}
\resizebox{1.\hsize}{!}{\includegraphics[angle=-90]{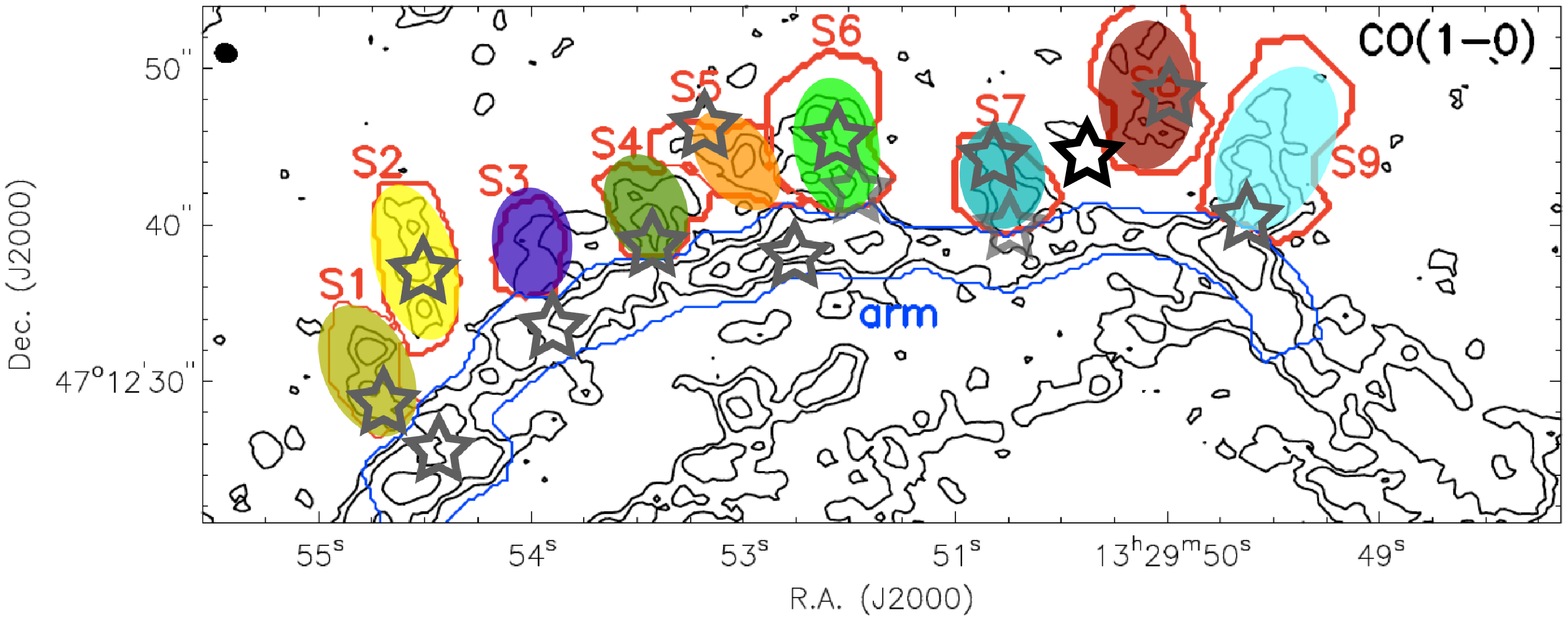}}
\end{center}
\caption{Relative age of star formation associated with our spurs 
(from young to old: dark blue -- cyan -- green -- yellow -- orange -- dark red shading)
based on different star formation tracers (see text for details). Preferred location of
star forming in each spur is marked by an open star symbol (dark grey). Star formation
outside spur S\,8 (black open star) and in the arm close to spurs (light open star) is
indicated as well. (See text for details.)
\label{fig:sf_age}}
\end{figure}

When using our crude classification for the age of the star formation site, we find no clear age trend across 
our nine spurs (color-coding in Fig.\,\ref{fig:sf_age} from blue (youngest) to red (oldest)). 

\subsubsection{Perpendicular to spiral arm = Along a spur}

The location of star formation sites relative to the spurs can be roughly classified into four categories: a) in 
the arm next to a spur base, b) at the base of the spur where it connects to the arm, c) in the middle of 
the spur, and d) at the tip of the spur. Spurs S\,1, S\,3 and S\,5 are next to star forming sites in the arm, 
while spurs S\,4 and S\,9 have most star formation occurring at their base. Prominent star formation in 
the middle of the spur occurs in spur S\,1, S\,2 and the almost dispersed spur S\,8. In spurs S\,5, S\,6 and 
S\,7 most star formation is found at the tip of the gas spur. In short, no obvious trend of the star formation 
location along a spur is found among our nine spurs (see open star symbols in Fig.\,\ref{fig:sf_age}).

When we combine the age classification with the location of star formation along a spur, we find no 
preferred location for a given age nor a trend along the spiral segment. However, it is interesting to 
note that some spurs (S\,6, S\,7, and S\,9) exhibit a clear age gradient along the spur with more recent
star formation activity being closest to the arm. For the remaining spurs, we see no clear spatial segregation
between different tracers for star formation activity and impact.

Taken together this suggests that star 
formation is not started in a preferred, fixed location relative to the spiral arm (in each spur). However,
there might be a preference for star formation onset more closely to the arm within each spur.
As star formation typically
proceeds for several Myr within an individual gas spur, this implies the onset mechanism has 
to act over a longer time-scale or the star formation process itself is not instantaneous but can be
sustained for a certain, few Myr long, time interval.

\subsection{Star Formation Feedback}
\label{subsec:feedback}
 
\begin{figure}
\includegraphics[width=140mm,angle=0]{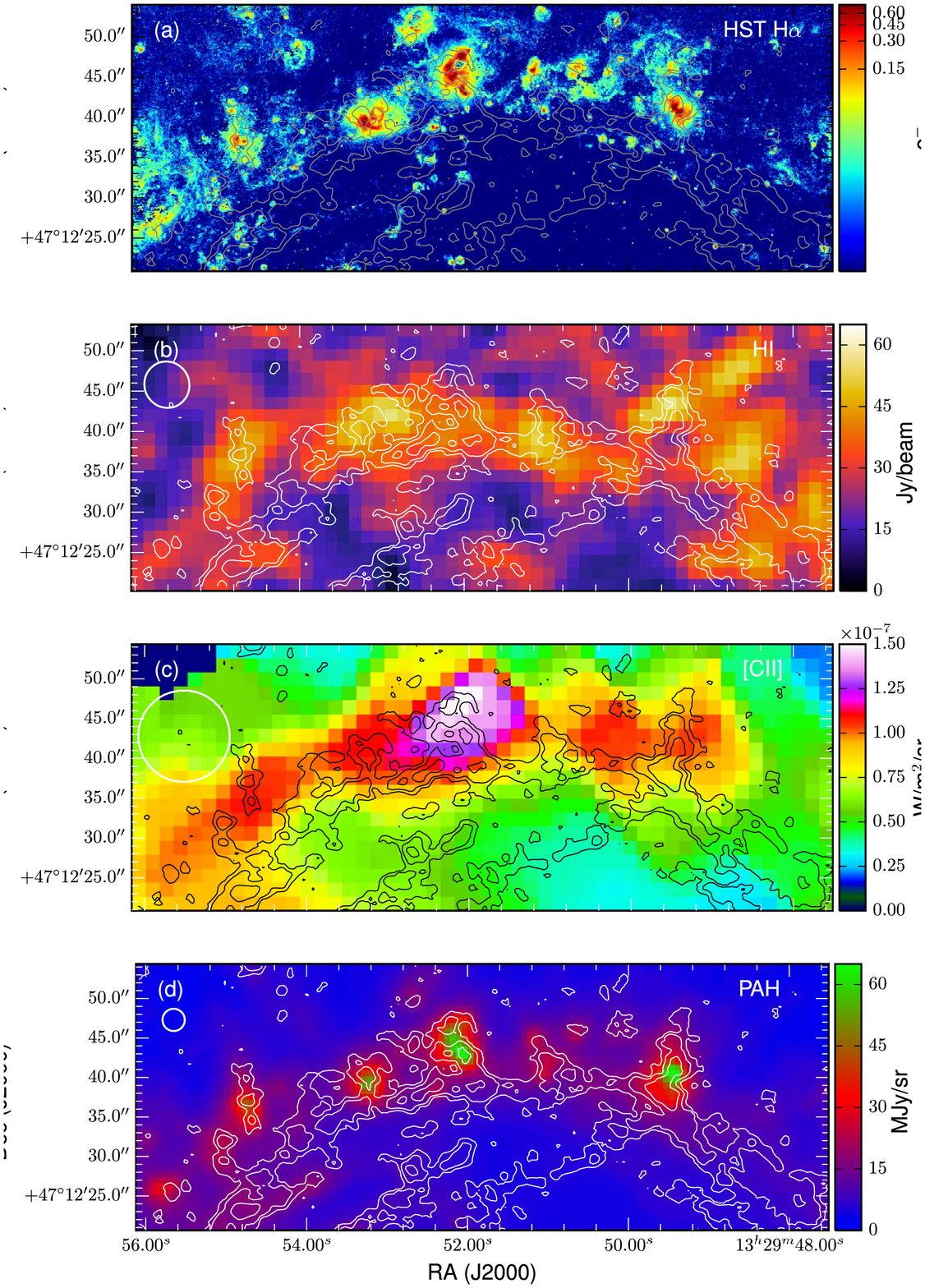}
\caption{Properties of the molecular gas and distribution of ISM tracers in the northern spiral arm of M51a:
(a) ionized hydrogen gas as traced by HST/ACS $H\alpha$ emission,
(b) atomic HI gas from VLA THINGS survey \citep{walter08},
(c) ionized carbon traced by [CII] line emission from PACS/Herschel \citep[e.g.][]{parkin13},
 and
(d) $\rm 8\,\mu m$ PAH emission from Spitzer IRAC imaging by SINGS \citep{kennicutt03}.
The contours in all images refer to the molecular gas surface density from Fig.\,\ref{fig:spur_co} (d) (thick lines only).
\label{fig:spur_ism}}
\end{figure}

To study the impact of star formation on the interstellar medium, we use the following data to
investigate the dissociation of molecular hydrogen (HI and $\rm H\alpha$) and the CO 
molecule ([CII] line at 158$\mu$m) as well as the heating of the interstellar dust ($\rm 8\mum$ emission corrected
for stellar contribution, and MIPS 24$\mum$ emission). Fig.\,\ref{fig:spur_ism} 
shows a comparison of some of these tracers to the molecular gas distribution as seen via its CO emission. 
We discuss the geometry and properties of the individual spurs below.

General trends of the impact of star formation onto the ISM can be summarized as follows: 

\begin{enumerate}[label=(\roman*)]

\item {\it The youngest stars heat the dust.} Hot dust/PAH emission along the spurs is always coincident
 with molecular gas (emission from CO)
(see Fig.\,\ref{fig:spur_ism}\,d), and the peaks in dust emission are consistent with the location 
of the youngest (i.e.\ below 3\,Myr) stellar clusters. Due to the lower resolution of the dust emission of 
$\sim$2'' compared to the HST imaging ($\sim$ 0.1'') for the young clusters, it is
difficult to search for small, but significant spatial offsets between the two tracers. However, in particular 
spurs S\,2 and S\,6 show some indication for an age differentiation even among the youngest stellar 
clusters (or at least their impact onto the surrounding ISM).

\item {\it The dissociation product of CO, [CII] emission line, can be observed  
after HII regions have ceased to exist.} A prime example is spur S\,8 where a bright [CII] emission peak (Fig.\,\ref{fig:spur_ism}\,c)
has no counterpart in H$\alpha$ emission (Fig.\,\ref{fig:spur_ism}\,a). Interestingly, the brightest [CII] and 
H$\alpha$ emission peaks (Fig.\,\ref{fig:spur_ism}\,a) do not show a 1-to-1 correspondence,
and the same is true for [CII] and hot dust/PAH emission. This suggests that the [CII] emission in the 
spurs tends to reach its peak brightness at a later point in time after the onset of star formation than the 
other two tracers and/or that [CII] emission has a longer decay time.

\item {\it The observed HI line emission can either be a $\rm H_2$ dissociation or 
$\rm H^+$ recombination product.} Given the location
of the HI peaks (Fig.\,\ref{fig:spur_ism}\,b), it seems that bright HI emission is more often observed as a recombination 
product, e.g.\ HI emission located downstream of spurs S\,5 and S\,8 with no associated HII emission, than a 
dissociation product, e.g.\ HI emission at base
of spur S\,7 associated with HII emission and the connection to the spiral arm. However, it is clear that for our arm regions, 
elevated HI emission is never observed upstream of the spiral arm, i.e.\ south of the molecular gas arm. This is consistent with 
the very high molecular gas fraction of $\rm > 80\%$ \citep{schuster07} that makes the need for $\rm H_2$ formation out of
atomic gas in/before the spiral arm shock obsolete. Note that \citet{tilanus89} already proposed that most of the HI
emission associated with the inner spiral arms of M\,51 is due to HI dissociation based on the observed offset between
the non-thermal radio continuum and the HI line emission. Our analysis suggests that the overall notion that HI emission
is due to the star formation process downstream of the spiral arm is correct, however, the exact identification of HI as being an
H$_2$ dissociation or an $\rm H^+$ recombination product is more subtle and will require more dedicated analysis.

\end{enumerate}

In the following we summarize the properties of each spur:

{\it Spur S\,1 --} 
The combined light of the HII regions and young stellar clusters associated with spur S\,1 are the
second faintest, their impact on the molecular ISM of the spur is not significant.  The H$\alpha$, HI and [CII]
line emissions show no evidence for enhancement at the location of this spur. As the hot dust/PAH
emission is slightly elevated, this suggests that some heating of the dust grains is happening.

{\it Spur S\,2 --}
Diffuse H$\alpha$ emission is coinciding with the entire extent of the molecular gas spur S\,2. The brightest
H$\alpha$ emission is located slightly east of the middle peak of the molecular distribution. The hot dust/PAH
emission is peaking between the two molecular gas emission peaks, just south of the H$\alpha$ peak. The brightest
HI emission in this spur is also found at the location of the hot dust/PAH peak. There is clearly enhanced
[CII] emission associated with this spur, however, due to the low resolution of the data it is difficult to draw any further
conclusions.

{\it Spur S\,3 --}
No emission from $\rm H\alpha$, HI, [CII] and/or $\rm 8\mum$ PAH/hot dust is associated with spur S\,3. This 
implies that there is no evidence for star formation activity in the recent past ($\rm \lessapprox 10-20\,Myr$). This is consistent with the absence
of HII regions and young stellar clusters and our interpretation of this spur as a potential site for future star formation.

{\it Spur S\,4 --}
The second brightest HII region (in our region under study) sits in spur S\,4, its impact onto the surrounding ISM 
is obvious from enhanced HI, [CII], and hot dust/PAH emission arising from this spur. The diffuse
H$\alpha$ emission surrounds the base of the spur and extends through a significant fraction of the molecular
spur towards its tip. It is interesting that a H$\alpha$ shell fits nicely into the kink at the western part of
the tip. The peak of the HI emission is offset from the prominent HII region towards the northwest and spur S\,5.
This could indicate that hydrogen is already recombining again in this region, as only one young stellar cluster and
no HII region is present in this area (see Fig.\,\ref{fig:spur_obj}). The hot dust/PAH emission coincides with the H$\alpha$ 
emission, while the [CII] emission has no distinct peak.

{\it Spur S\,5 --}
The H$\alpha$ emission in the tip of spur S5 has no prominent counterparts in the other tracers. This is
not surprisingly, as both the HII regions and the young stellar cluster are the smallest of all the ones that are
hosted in spurs. Thus no strong
impact is necessarily expected. In addition, due to the proximity of the very prominent star formation in spur S\,6
and the low angular resolution of most tracers it is difficult to uniquely associate them with spur S\,5. 

{\it Spur S\,6 --}
The brightest HII region and the largest number of young stellar clusters (including the most massive one) are found
in spur S\,6, thus a strong impact of the star formation onto the surrounding ISM is expected.
A chain of HII regions is located along the hook-shaped gas spur. The regions in the south are surrounded
by less diffuse H$\alpha$ emission and are coincident with molecular gas and hot dust/PAH emission. 
The HII regions in the middle and the far north straddle the molecular gas emission peak of the spur. These regions
are also accompanied with most of the young stellar clusters, while most of the older ($\rm 3\,Myr < {\it t} \le 10\,Myr$)
clusters are found in the ridge between the molecular gas emission peaks at the base and the tip of the spur. The brightest
peak in [CII] emission is associated with spur S\,6 and the peak within this brightest [CII] emission is found at the tip
of the spur. Interestingly, no enhanced HI emission is seen from the tip of the spur, but it is rather seen
from the base and the neighboring arm segment. 

{\it Spur S\,7 --}
The situation for spur S\,7 is more complex, the HII region at the base of the spur is associated with a peak in HI and 
$\rm 8\mum$ hot dust/PAH emission while the two HII regions at the tip of the spur are only co-spatial with
enhanced non-stellar $\rm 8\mum$ emission. There is no direct evidence for [CII] emission arising 
from spur S7 at all given the resolution of the data. This behavior could indicate that the impact of star formation
is stronger close to the arm (given that the H$\alpha$ luminosity of this region is only about 10\% of the 
one from the regions at the tip) and/or that the HII regions at the tip are older given their larger sizes.

{\it Spur S\,8 --}
The [CII] emission shows a clear peak located roughly between spur S\,8 and the neighboring arm segment.
Interestingly, the young stellar clusters are on the eastern side of this peak and no HII region is associated
with this [CII] peak. However, faint, diffuse H$\alpha$ emission is visible in the HST imaging  of this region,
while no HI peak is visible either. This suggests that the nearby young stellar clusters are providing enough
energy input to dissociate CO. A similar [CII]/H$\alpha$/young stellar cluster geometry can be found further 
along this spiral arm at 13:29:46.9 +47:12:29 (J2000) outside our region studied in detail. Interestingly the 
gas spur S\,8 itself coincides with HII regions, lightly
enhanced HI and dust/PAH emission, implying that star formation is still impacting all components of the ISM.

{\it Spur S\,9 --}
The H$\alpha$ emission associated with the very bright HII region at the base of spur S\,9 shows a bi-conical morphology
that is basically oriented perpendicular to the extent of the spur. The peak of the hot dust/PAH emission coincides
with the southern HII region and the location of the very young stellar clusters. However, the hot dust/PAH emission
appears to be 
a slightly shifted towards the arm. Given the lower resolution of the IRAC data, we can not exclude that this is an artifact.
The northern HII region within this spur sits
in a kink of the molecular gas emission distribution close to the spur's tip. Strong HI emission is arising from this region.
The [CII] emission peaks in the northern half of this spur, but it is about 25-30\% less luminous in surface brightness
(or about 50\% in integrated flux) than the emission associated with spur S\,6, roughly consistent with the difference
in H$\alpha$ luminosity between the HII regions of these two spurs. 

Given the fact that star formation is concentrated towards the gas spurs one could conceive that the star formation
happening in a spur could trigger more star formation events. We use our spiral arm segment to search for signs 
of star formation activity triggered by stellar feedback. Given our 
resolution and tracers used, we find that spur S\,6 is the only spur showing potential signs of triggered star 
formation activity among our nine spurs studied -- based on its particular morphology.

Spur S\,6 is not exactly oriented perpendicular to the gas arm, but appears to be tilted more eastward. In addition
the low level CO emission exhibits a hook-like appearance at the tip of the spur. Two young
stellar clusters with an age of $\rm 3-10\,Myr$ \citep{chandar11,chandar16} sit within this hook. While a large number of
$\rm \le 3\,Myr$ young stellar clusters are distributed along the hook continuing along the Western side of the spur
toward the spiral arm (see Fig.\,\ref{fig:spur_obj}). Similarly, strong H$\alpha$ emission is  
associated with the gas hook and the Western side of the gas spur, while more diffuse, low level H$\alpha$
emission is seen inside the hook. This geometry can suggest that star formation first occurred at the location 
inside the hook and that due to stellar feedback onto the surrounding gas more star formation has been triggered.
This could explain part of the tentative age gradient present in this region. 

While it is not clear that stellar feedback has indeed triggered star formation within the spurs, it seems clear from
the molecular (CO) and ionized (H$\alpha$) gas morphology that star formation feedback is impacting the
molecular gas distribution, e.g.\ spurs S\,4 and S\,9.


\section{Discussion}
\label{sec:discussion}

We discuss our results in the context of a simple spiral density wave picture which predicts a clear offset signature
between spiral arm and star forming sites (\S \ref{subsec:sdw}). Given the surprisingly close relation between gas
spurs and star forming sites, we compare our findings to expectations from simulations (\S \ref{subsec:spurs}).

\subsection{A simple spiral density wave picture and its implications}
\label{subsec:sdw}

In the framework of a spiral density wave the following picture can be put forward: At the location of
the spiral arm potential the gas is efficiently collected, compressed and starts to 
collapse and form stars. In this simple picture, one would expect that the collapse of the gas clumps 
and the subsequent star formation is always taking place in the same location, i.e.\ at or close to the spiral
arm potential. In the case of a spiral density wave, the spiral arm rotates at a constant pattern speed. Thus
the differential disk rotation will cause a constantly varying offset between the gas spiral and star forming regions 
(that decoupled from the gas motion) due to the difference in age. Note that this picture makes no assumptions
about the exact cause of cloud formation or their collapse. 

\begin{figure}[ht]
\begin{center}
\resizebox{.9\hsize}{!}{\includegraphics[angle=0]{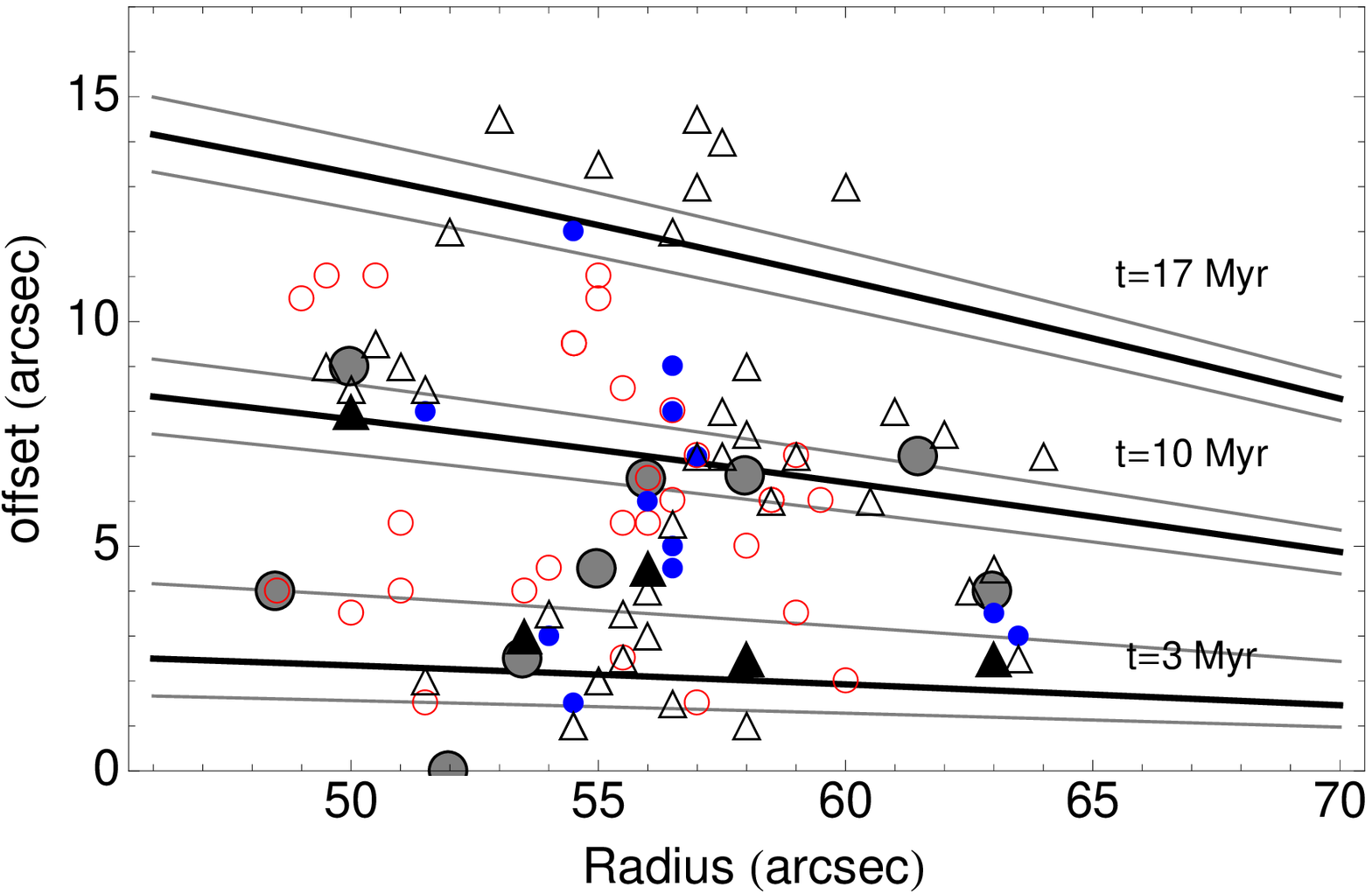}}
\end{center}
\caption{Offset of star forming regions from gas spiral arm as function of galacto-centric position of the arm. The offset was determined
(by eye) in perpendicular direction to the gas arm in a deprojected image; the typical error on both parameters
is $\sim$0.2'' due to the uncertainty in determining the ridge line of the gas arm. Small shifts in radius have been
applied for better readability of the plot. Small filled circles show young stellar clusters from \citet{chandar16} and the 
color coding corresponds to young stellar cluster ages of 
$\rm log({\it t}(yr)) < 6.5$ (blue filled), and
$\rm 6.5 \le log({\it t}(yr)) < 7.0$ (red open). 
The open and filled
black triangles mark the locations of individual HII regions and HII complexes from \citet{lee11}. 24$\mu$m peak positions are given as large grey filled
circle. The thick solid lines correspond to the expected offsets from the spiral arm after 3, 10 and 17\,Myr ($\pm$1\,Myr). The offsets
perpendicular to the arm \citep[with a given pitch angle of $\rm {\it i}_p=20^o$;][]{patrikeev06} are based on a simple picture in 
which star formation occurs instantaneously with the passage of the spiral arm modeled as a simple kinematic wave with
a pattern speed of $\rm \Omega_{P,spiral}=53\,km\,s^{-1}kpc^{-1}$  \citep[][]{querejeta16} and assuming the rotation curve from \citet{meidt13a}. 
No clear trend is evident for all or individual star formation tracers.
\label{fig:offset}}
\end{figure}

Our results from section \S \ref{sec:arm} and \ref{sec:process} are already qualitatively in disagreement with this picture.
However, a certain stochasticity in the star formation onset within GMCs could introduce some scatter. Therefore we
measure the deprojected radial offset of different star formation tracers from their neighboring spiral arm location, namely
for the 24$\mu$m peaks, the individual HII regions and complexes from \citet{lee11}, and the young ($\rm <10\,Myr$)
stellar clusters \citep[separated into younger and older than 3\,Myr;][]{chandar16} and compare it to the expected offset
for a constant pattern angular speed of $\rm \Omega_{P,spiral}=53\,kms^{-1}kpc^{-1}$ \citep{querejeta16} of a spiral arm with
a fixed pitch angle of $\rm {\it i}_p=20^o$ \citep{patrikeev06}. In Fig.\,\ref{fig:offset}
no clear trend in the offset perpendicular from the arm is obvious for stellar clusters younger (older) than 3\,Myr. Similarly individual 
HII regions show a wide
spread in that offset corresponding to times of up to 8\,Myr (clearly larger than a HII region lifetime). A trend for shorter separation
times is implied when concentrating on the brightest star forming regions as evidenced by the 24$\mu$m peaks and the HII complexes,
however, no preference for a single separation time is evident. Thus we conclude that our qualitative picture is correct.
The varying offset between the young star forming regions and the gas spiral arm implies that the most massive star 
formation is not always starting in exactly the same location relative to the spiral arm potential. 

Fig.\,\ref{fig:offset} paints a complicated picture of star formation in the spiral arm.Ê Most of the young regions nearest to the molecular arm 
ridge are consistent with forming in the spiral arm, including the embedded star formations sites traced by 24$\mu$m and the HII region 
complexes. However, the young regions furthest from the arm (including some 3\,Myr and10\,Myr old stellar clusters and HII regions)
are beyond where they could be in the case of instantaneous star formation in a simple propagating kinematic wave.ÊThese star forming 
regions do indeed appear to have formed very near to their present location, i.e.\ at the location of the spurs.ÊOnly if there are positive radial 
and azimuthal, i.e.\ outward, flows (on the order of $\rm 10-15\,km\,s^{-1}$), the location of these star forming sites might still be consistent 
with forming in the spiral arm, as these (additional) radial motions would allow young 
star formation sites to move much further away from the spiral ridge than illustrated in Fig.\,\ref{fig:offset}. However, the presences of such 
outward radial flows are particularly unlikely.Ê The observed gas kinematics imply strongly radially inward motions already at the location of
the gaseous spiral arm \citep[see][]{meidt13a}. 

The star formation sites furthest away from the arm might alternatively arrive at their present location, if their progenitor clouds formed in the spiral arm, but the 
onset of star formation has been delayed.Ê
Adding an additional 5-8\,Myr before star formation occurs would allow the $\sim$3\,Myr 
old stellar clusters and HII regions to reach their present positions.ÊThis timescale is consistent with the crossing time for the observed
spiral arm and spur clouds, i.e.\ $\rm {\it t}_{cross} = {\it r}_{cloud}/\Delta\,{\it v}_{cloud}$ (with $\rm {\it r}_{cloud}\approx 40\,pc$ and
$\rm  \Delta\,{\it v}_{cloud,arm}\approx 8\,km\,s^{-1}$ and $\rm \Delta\,{\it v}_{cloud,spur}\approx 5\,km\,s^{-1}$ resulting in $\rm 
{\it t}_{cross,arm}\approx 5\,Myr$ and $\rm {\it t}_{cross,spur}\approx 8\,Myr$ for spiral arm and spur clouds, respectively).Ê At fixed size, low mass clouds with smaller velocity dispersions and longer crossing times would appear further from the spiral arm ridge than their higher-mass counterparts.Ê Note, though, that not all observed regions (particularly those with smallest offset from the arm) would require such a delay to be consistent with the spiral arm formation scenario. Ê

Alternatively, we can consider a scenario where star formation is initiated within the spiral arm but yields new stars with a delay to explain
our observation of star formation occurring within spurs. In this case the delay represents the time to form the spur itself, i.e.\ from a sheared 
arm cloud.Ê With the measured offsets it is not possible to distinguish between spur formation through the evolution of spiral arm clouds 
\citep[e.g.][]{dobbs08,dobbs13} or the case in which the spurs are independently evolving structures forming via gravitational instability, 
such as envisaged by \citet{kim02}.ÊThe lack of a 
clear gradient in the star formation ages across the spurs appears to be inconsistent with other suggested formation mechanisms 
\citep[e.g.][see also below]{renaud13,wada11}.Ê The lack of a clear age gradient also disfavors star formation happening solely in the 
spiral arm (at least in the simple 
propagating wave picture).Ê We thus conclude that many of the observed star formation sites must be genuinely associated with the 
spurs, rather than the spiral arm, whereas others are consistent with forming in the spiral arm.Ê More generally, young regions at the 
observed offsets cannot be ascribed a single formation mechanism or a single characteristic timescale. Ê

From this analysis we can conclude that star formation proceeds with a variety of timescales in/near the spiral arms.Ê Despite the observed 
complexity in the positions of the sites of recent star formation relative to the spurs, the evidence is consistent with a mixture of mechanisms 
that lead to both coarse- and fine-tuning of the star formation timescale.Ê Broadly, we identify star formation occurring in two main modes: 
star formation within the spiral arms and star formation starting independently within spurs, presumably when the clouds created in the 
process of spur formation are sufficiently massive and bound enough to collapse and form stars.Ê This leads generally to two characteristic 
zones for observed star forming regions relative to the spiral arm at (roughly) fixed age: near and far from the spiral arm ridge.Ê A spreading 
throughout these zones is the result of additional fine-tuning in the timescale, determined by properties of the individual clouds themselves.Ê 
We find that the scatter observed around the arm and spur zones is qualitatively consistent with the additional dependence of the star 
formation timescale specifically on the crossing time of the cloud (see above). Ê

Although the formation of individual spurs and the star-forming clouds within them are likely subject to local conditions, we speculate that 
spur, and subsequent star, formation depends on gas dynamics on scales larger than molecular clouds and that the process may even arise 
with material processed independently of the arm cloud population.Ê Our comparison of the CO distribution imaged at different resolutions 
suggests that more diffuse molecular gas may be distributed with a regularity reminiscent of the apparent regularity in the spur population.Ê 
Spur formation in this case might occur through the compression of diffuse gas and subsequent gravitational instability as envisaged by 
\citet{kim02}, rather than as the result of clouds from the arm shearing out as they exit the arm and pass in to the interarm.Ê In the former 
scenario, a new population of clouds would form as part of the spurs, independently of the spiral arm cloud population.Ê More critically, the 
process of star formation in spurs would be decoupled from star formation in the spiral arms. ÊÊ

The fact that spurs can support star formation independently of star formation starting in the spiral arm would have important implications for 
global gas consumption within and among galaxies.Ê It has been suggested that, in some instances, the gas kinematic characteristic of flow 
through a spiral arm perturbation may lead to a suppression of star formation, i.e.\ due to enhanced turbulence in the spiral shock 
\citep[e.g.][]{kim02} or as a result of cloud-cloud collisions \citep[e.g.][]{dobbs08} or under the influence of dynamical pressure 
\citep{meidt13a}.Ê But even when the spiral arm suppresses star formation, the overall dynamics of the spiral can still lead to (at least) modest 
rates of star formation by promoting spur formation.Ê Such spur-based star formation would then be responsible for the low level of star 
formation observed in M51 in the region of the spiral arm where star formation is suppressed \citep[relative to the high rate expected given 
the observed gas surface density][]{meidt13a} but where spur formation appears to continue successfully.Ê The region with lowered star 
formation in the spiral under study here directly connects to our segment analyzed at smaller radii.
Thus, even when gas kinematics leads to a suppression of 
star formation internal to the arm, overall spiral arm dynamics could still provide the avenue for star formation through the creation of spurs. Ê

Spur-based star formation would also lead to localized pockets of recent star formation and groupings of young stellar clusters.Ê This can make it 
difficult to successfully use offsets to measure spiral pattern speeds.
A direct implication of our analysis is then that the observed offset between (gas) spiral arms and young star formation has a significant
intrinsic scatter. Thus its use to determine the pattern speed of spiral arms \citep[e.g.][]{egusa09} will result in larger uncertainties
or less clear answers than naively expected. The significant variation in star formation age and location will also affect other
applications that use this simple picture and lead to less clear signatures. This might explain some of the conflicting findings
reported in the literature \citep[e.g.][]{tamburro08,foyle10}, especially when taking into account that the interpretation of the gas
properties are also affected by dissociation and recombination timescales.

\subsection{Formation and Evolution of Gas Spurs}
\label{subsec:spurs}

Theoretical models and simulations \citep[e.g.][]{kim06,kim02,dobbs08,dobbs11} developed the following 
picture for the formation of GMCs and sub-sequent star formation in these dense gas complexes:
The formation of gas peaks inside the gas spiral arm can be due to agglomeration of small clouds 
\citep{dobbs08} and/or self-gravity \citep{dobbs08,tan00,kim02}, or the magneto-Jeans instability \citep{kim02}. 
These overdensities will become gravitationally unstable and 
fragment reaching cloud masses up to $\rm 10^6\,\msun$. Due to shear (induced by the spiral 
potential) these gas fragments are stretched 
perpendicular to the spiral. Large GMCs can occur at preferred locations within the spiral arms with a regular spacing 
given by the Jeans length. Or alternatively with the agglomeration scenario there is a quasi-periodic spacing associated 
with the epicyclic frequency. The gas spurs are the result of these
GMC overdensities becoming stretched out.

Our observations support this scenario only partially as the gas spiral arm appears fairly smooth (with variations
within a factor of 2-3 in brightness) and
shows no preferred distance between CO peaks at full ($\sim$ 40\,pc) resolution nor the location of the identified 
GMCs (see Fig.\,\ref{fig:spur_obj}a). However, at a 
lower resolution of 3.0$\,\as$ ($\sim$ 110\,pc)
CO peaks with roughly regular spacing become evident. This implies that more diffuse gas on spatial scales larger 
than typical GMCs ($\rm \sim 40\,pc \approx 1\,\as$ in M51) is organized in a more regular pattern. As
already seen with the feathers \citep[dust lanes emanating from the spiral arm, e.g.][]{lavigne06} the 
gas spurs closely represent those features seen in simulations. We find no significant difference in 
the properties of GMCs located in the arm or the spurs, suggesting that no large transformation of
more bound structures is happening during the transition from arm to spur. The lower fraction of diffuse emission 
in the spurs could point to the fact that spurs mark the location of most efficient compression/assembly of gas into bound structures.
It is interesting to note that the spurs contain indeed GMCs with masses similar to those found by \citet{kim06}. 

The most interesting observation is that massive clustered star formation seems to be almost entirely
associated with gas spurs. This immediately implies that stellar feedback should have a significant
impact on the shape and evolution of these gas spurs. Our analysis is inconclusive regarding the location
where the onset of star formation occurs. 
Taken all results together we find evidence for a star formation onset with no specific preference for a position
along a spur and an apparent avoidance of star formation starting within the spiral arm itself. Thus it is not
consistent with the simple assumption of a gas density threshold above which star formation starts, as there is no
(large) difference in the gas and GMC properties between spiral arm and spur GMCs. The most obvious trend appears to
be that the more gas-rich a spur gets the higher its level of star formation activity is (e.g.\ S\,6 \& S\,9), and spur S\,3 being the 
the least gas-rich one showing no sign of ongoing star formation activity.

Our detailed high-resolution analysis also shows that the use of lower resolution imaging could be misleading, as several
star formation events (separated in age) can be present within a single spur and only the brightest event would dominate
the light at different spurs. For example, \citet{elmegreen14} analyzed $\sim 2\,\as$ resolution 3.6$\mu$m, H$\alpha$ and SDSS
images to identify the youngest star forming sites along spiral arms in five nearby spiral galaxies including M\,51. Their embedded
sources 1 and 2 correspond to our spurs S\,2 and S\,6. It is clear that star formation has been proceeding in S\,6 for quite a while
(see \S \ref{subsec:gas_sf})
including evidence for star formation induced by stellar feedback (see \S \ref{subsec:feedback}). Therefore it seems that analysis and 
interpretation need to account for the presence of multiple star formation events or a prolonged period of star formation even 
for large complexes. Our derived average GMC masses in spurs S\,2 and S\,6 are at the lower ($\sim 1\times10^6\,\msun$) 
and higher ($\sim 4\times10^6\,\msun$) end of gas
masses observed. However, in any case they are well below the $\rm 10^7\,\msun$ inferred by \citet{elmegreen14} for these regions.

In order to
infer an estimate of the star formation efficiency (SFE) we compare the average mass in GMCs to that in young stellar clusters.
(Note that our estimate for the cluster mass is most likely a lower limit as stellar clusters lose already a significant fraction (a few 10 percent depending on the assumptions) of their mass within their first 10\,Myr.)
We find that spurs S\,1, S\,4, S\,5, S\,6, S\,8, and the arm have a $\rm SFE<1\%$ while SFE is more than ten times higher in 
spurs S\,2, S\,7, and S\,9 (spur S\,3 has no stellar clusters and is excluded). These SFEs are low compared to values derived for 
Galactic GMCs of a few percents and in particular for cloud regions with observed clustered star formation where SFEs of a few 10\%
have been derived \citep[e.g.\ review by][]{padoan14}. The low SFEs could mean
that some of the lower mass clouds are not collapsing resulting in apparently lower SFE, that some of the gas associated with GMCs by the 
identification algorithm \citep[see appendix of][for details]{colombo14a} is not bound, again causing a lower SFE, or that the young
stellar clusters have already experienced a much more significant mass loss than assumed. A large population of stellar clusters with
masses below our detection limit would result in even lower SFE values, while very young clusters residing in HII regions could potentially
be more massive and lead to potentially higher SFEs. In any case the large variation in derived
SFE suggests that star formation is not proceeding uniformly across our nine spurs considered.

In the turbulent picture one would expect that higher internal turbulence in GMCs leads to higher star formation rates as more gas
can be pushed to higher gas densities suitable for stars to form. Our spur GMCs exhibit slightly lower line widths than arm GMCs 
and the arm GMCs have abundant star formation associated with them whereas no much star formation activity is observed in
the arm GMCs -- contrary to the simple expectation from the turbulent picture. This supports our interpretation that (massive) star 
formation in the spiral arm is significantly lowered or not occurring on relevant levels. 
\citet{meidt13a} proposed that GMCs in spiral arms
might be stabilized through dynamic pressure increased by the streaming motions present in the spiral arms. This scenario
could explain simultaneously the lack or shortage of star formation occurring in the spiral arm and the (slightly) lower observed velocity
dispersion in the spurs. Other possibilities could be enhanced turbulence in the spiral shock \citep{kim02} or increased cloud-cloud
collisions preventing immediate cloud collapse.
In any case, this would imply that spiral galaxies with less strong spiral potential should show a pattern that starts
to deviate from M\,51's strong separation of star formation sites and gas spiral arms, independent of the inferred spiral pattern
speed. Thus we interpret the apparent time delay between the spiral arm and the location of 
star formation being due to the time it takes to form gravitationally bound structures within GMCs rather than a delay between the 
presence of such structures and the actual onset of star formation within them. 

Recently, \citet{renaud13} proposed a different formation mechanism for gas spurs as the one described above, 
namely via Kevin-Helmholtz instabilities. The simulated region shown \citet[see Fig.\,13 of][]{renaud13} roughly resembles the
geometry of our region (orientation of the spiral arm with respect to the
galaxy center and galactic rotation). The spurs in the simulation have an age gradient in the sense that they start to 
dissolve at shorter galactic radii while they are still forming at larger galactic radii.
In this scenario, one could expect to see an age gradient for star forming sites 
across neighboring spurs. Our analysis finds no evidence for such an age gradient, implying that the proposed picture
is too simplistic or not fully applicable.

In any case, we conclude that there seems to be a close connection between spurs and massive cluster formation in M51 which
suggests that spurs might be a requirement for the existence of massive clustered star formation. 
Thus the mechanism for or cause of gas spur formation is a pre-requisite to form (super-)giant 
HII regions and complexes of young stellar clusters. A large statistical sample of arm/spurs GMC and star formation properties is 
required to properly address cloud and star formation in spurs.


\section{Summary and Conclusion}
\label{sec:summary}

In order to better understand the star formation process along spiral arms, we combined high quality and high spatial
resolution observations of the interstellar medium and tracers of recent star formation for a spiral arm segment in the
disk of the nearby grand-design spiral galaxy M\,51. The selected arm region is consistent with being driven by a spiral
density wave, in the sense that the spiral arm is significantly contributing to the gravitational potential. 

Our analysis shows that the picture is more complex than inferred from the simple picture where star formation is started
inside a gas spiral arm (by whatever physical process). While a close connection between gas spurs and massive star formation
is observed, making a causal connection is difficult. The impact of the recent star formation on the gas spurs is evident in different
ISM tracers. In particular we find that:

\begin{itemize}

\item
The molecular gas in the selected spiral arm region is distributed into a distinct arm from which gas spurs
emanate in an almost perpendicular direction. Detailed analysis shows that Giant Molecular Associations (GMAs)
are caused by blending of gas spurs with their neighboring arm segment and are therefore not single 
entities (see \S\,\ref{subsec:co_gmas}). While the overall gas surface density in the spurs is lower than in the arm itself
it appears that the gas in the spurs is on average more bound, as the fraction of gas in GMCs is higher in spurs.
No other significant differences in properties of GMCs located in the arm or spurs are found.

\item
Star formation activity is strongly biased towards the spurs, with only a few star forming sites located inside the
spiral arm. No trend in the age of the star formation events is seen either between spurs or along individual spurs. Together 
with the tendency for massive star formation to occur at a preferred location along spurs, this suggests that the star formation onset
is not solely set within or close to the spiral arm. Other stabilizing processes might inhibit the onset of star formation or
prolong the collapse of clouds. In addition, rapid dispersal of stars formed in clusters might play a role as well.

\item
Comparison of the location of emission from heated dust, atomic, and ionized gas reveals that star formation feedback is mostly confined
to the region downstream from the spiral arm and often at the tip of the gas spurs. We speculate that the star formation in the upper half
of spur S\,6 is triggered by stellar feedback given its peculiar shape. Atomic hydrogen emission seems to be due to either $\rm H_2$
dissociation or recombination from the ionized gas. We also identify a regions of bright [CII] emission without associated H$\alpha$
emission but several young stellar clusters which suggests that [CII] emission is powered over a longer timescale than H$\alpha$.

\item
Our detailed analysis suggests that the offset between star formation sites and a gas/dust spiral arm cannot be explained by simple rotation
of the spiral arm pattern, as star formation appears not to start at similar locations in the spiral arm. Thus interpretations relying on the simple assumption
that star formation is started in a single (fixed) location, i.e.\ the gas arm, can lead to incorrect or inconclusive results. Further this implies that models that predict
star formation onset solely in the spiral are too simplistic and need to take into account additional mechanisms that could inhibit or prolong 
immediate cloud collapse. Possible candidates are the increased dynamic pressure due to streaming motions in the spiral arms or
stabilization due to magnetic fields. 
We speculate that the offset between star forming sites and gas arms might be more a function of the strength of the spiral arm potential than the actual pattern spiral speed.

\end{itemize}

Based on our results we conclude that analysis of a large statistical sample of spurs in galaxies that host differing spiral arm potentials
will be required to provide the insights to make significant progress in our understanding of the role of spiral arms for star formation.


\acknowledgments
We thank the IRAM staff for their support during the observations with
the Plateau de Bure interferometer and the 30m telescope. 
SEM and MQ acknowledge funding from the Deutsche Forschungsgemeinschaft (DFG) via grant SCHI 536/7-2 as part of the priority program SPP 1573 'ISM-SPP: Physics of the Interstellar Medium'.
CLD acknowledges funding from the European Research Council for the FP7 ERC starting grant project LOCALSTAR.
JP acknowledges support from the CNRS programme \`{P}hysique et Chimie du Milieu Interstellair\'{e} (PCMI). 
MQ acknowledges the International Max Planck Research School for Astronomy and Cosmic Physics at the University of Heidelberg (IMPRS-HD). S.G.B. thanks support from Spanish grant AYA2012-32295. 
We acknowledge financial support to the DAGAL network from the People Programme (Marie Curie Actions) of the European UnionÕs Seventh Framework Programme FP7/2007- 2013/ under REA grant agreement number PITN-GA-2011-289313.
ES thank NRAO for their support and hospitality during her visits in Socorro.
ES thanks the Kavli Institute for Theoretical Physics for hospitality during the writing of this paper.
IRAM is supported by INSU/CNRS (France), MPG (Germany) and IGN (Spain). 

{\it Facilities:} \facility{IRAM (PdBI)}, \facility{IRAM (30m)}, \facility{HST (ACS)}, \facility{HST (NICMOS)}, \facility{GALEX}, \facility{NRAO (VLA)}, \facility{Herschel (PACS)}, \facility{Spitzer (IRAC)}, \facility{Spitzer (MIPS)}.


\clearpage

\bibliography{spurs-ref}


\clearpage


\footnotesize
\begin{deluxetable}{crrrrrrrrrrr}
\tabletypesize{\footnotesize}
\tablecaption{Molecular Gas Properties of the Northern Spiral Segment}
\tablehead{
\colhead{Spur} &
\colhead{$\rm {\it I}_{CO}$} &
\colhead{$\rm {\it S}_{CO}$} &
\colhead{$\rm {\it M}_{H_2}$} &
\colhead{Area} &
\colhead{$\rm \overline{\Sigma _{{\it M}_{H_2}}}$} &
\colhead{$\rm \langle\Sigma _{{\it M}_{H_2}}\rangle$} &
\colhead{$\rm \Sigma_{{\it M}_{H_2}}^{max}$} &
\colhead{$\rm \overline{{\it T}_{CO}}$} &
\colhead{$\rm \langle{\it T}_{CO}\rangle$} &
\colhead{$\rm {\it T}_{CO}^{max}$} &
\colhead{$\rm \overline{\sigma}$}
\\
\colhead{\#} &
\colhead{($\rm K\,km\,s^{-1}$)} &
\colhead{($\rm Jy\,km\,s^{-1}$)} &
\colhead{($\rm 10^6\,M_{\odot}$)} &
\colhead{($\rm 10^3\,pc^2$)} &
\colhead{($\rm M_{\odot}\,pc^{-2}$)} &
\colhead{($\rm M_{\odot}\,pc^{-2}$)} &
\colhead{($\rm M_{\odot}\,pc^{-2}$)} &
\colhead{(K)}&
\colhead{(K)}&
\colhead{(K)} &
\colhead{(km/s)}
}
\startdata
1   &   870 &   10.7  &    4.8  &   47.5   &  100  & 80 & 360 & 2.2 & 2.0 & 5.7 & 5.6\\
2   &   930 &   11.3  &    5.1  &   63.9   &   80  &  60 & 290 & 2.1 & 1.7 & 6.1 & 4.9\\
3   &   470 &    5.8  &    2.6  &   31.9   &   80  &   80 & 220 & 2.0 & 2.0 & 4.0 & 5.6\\
4   &   920 &   11.2  &    5.1  &   43.1   &  120  & 110 & 300 & 2.7 & 2.6 & 5.9 & 5.8\\
5   &   760 &    9.2  &    4.2  &   42.6   &  100  &  90 & 320 & 2.4 & 2.1 & 6.2 & 5.5\\
6   &  2200 &   26.8  &   12.1  &   84.8   &  140  &110 & 480 & 3.1 & 2.6 & 8.5 & 5.5\\
7   &   980 &   11.9  &    5.4  &   44.5   &  120  &  110 & 320 & 2.7 &2.5 & 6.2 & 5.7\\
8   &   790 &    9.6  &    4.4  &   85.0   &   50  &   40  & 220 & 1.9 & 1.8 & 5.3 & 4.1\\
9   &  2200 &   26.4  &   12.0  &  111.9   &  110  & 70& 510 & 2.6 & 2.1 & 8.3 & 5.6\\ \hline
mean & 1120 & 13.7 & 6.2 & 61.7 & 100& 80& 340 & 2.4 & 2.2 & 6.2 & 5.4\\ \hline
arm & 15000 &    182  &   82.6  &  419.7   &  200  & 180 & 630 & 3.0 & 2.9 & 7.8 & 8.1\\
\enddata
\tablecomments{The notation of the spurs and the arm as well as their areas are indicated in 
Fig.\,\ref{fig:region}. Values 
are derived assuming a distance to M51a of 7.6\,Mpc and a Galactic conversion factor of $\rm 
2x10^{20}\,\,cm^{-2}K^{-1}km^{-1}s$ for CO intensity into \htwo ~gas mass. For the average \htwo ~gas surface density
we list both the mean $\rm \overline{\Sigma _{{\it M}_{H_2}}}$ and median $\rm \langle\Sigma_{{\it M}_{H_2}}\rangle$
values. For the average CO peak brightness temperature we list both the mean $\rm \overline{{\it T}_{CO}}$
and median $\rm \langle{\it T}_{CO}\rangle$ values.
\label{tab:co}}
\end{deluxetable}
\normalsize

\clearpage


\begin{deluxetable}{c|crrrrrr}
\tabletypesize{\normalsize}
\tablecaption{Properties of GMCs located in Northern Spiral Segment}
\tablehead{
\colhead{Spur} &
\multicolumn{7}{c}{GMC}
\\
\colhead{ID} &
\colhead{\#}&
\colhead{\it r} &
\colhead{$\rm \Delta\,{\it v}$} &
\colhead{$\rm {\it M}_{H_2}$} &
\colhead{$\rm {\it M}_{H_2}/{\it A}$} &
\colhead{$\rm {\it M}_{H_2,GMC}$} &
\colhead{$\rm \frac{{\it M}_{H_2,GMC}}{{\it M}_{H_2,tot}}$}
\\
& &
\colhead{(pc)} &
\colhead{($\rm km\,s^{-1}$)} &
\colhead{($\rm 10^5\,M_{\odot}$)} &
\colhead{($\rm M_{\odot}/pc^2$)} &
\colhead{($\rm 10^6\,M_{\odot}$)} &
}
\startdata
1 & 4 & 38 & 7.8 & 11.0 & 93 & 4.4 & 0.92\\
2 &   3 &  38 & 5.6 &  9.8 &  46 & 3.0 & 0.58 \\    
3 &   3 &  48 & 5.8 & 10.3 &  97 & 3.1 & 1.19 \\
4 &   1 & 120 & 4.0 & 61.5 & 143 &  6.2 & 1.20\\    
5 &   1 &  60 & 4.4 & 29.9 &  69 & 3.0 & 0.71 \\ 
6 &   3 &  62 & 7.4 & 42.9 & 152 & 12.9 & 1.06 \\    
7 &   3 &  55 & 4.9 & 25.6 & 173 & 7.7 & 1.42\\    
8 &   1 & 125 & 8.2 & 62.2 &  73 & 6.2 & 0.65 \\    
9 &   6 &  45 & 5.6 & 14.5 &  78 & 8.7 & 0.73 \\ \hline
mean & 2.8 &  66 & 6.0 & 29.7 &  103 & 6.1 & 0.94\\ \hline
arm & 27  &  51 & 7.2 & 23.9 & 150 & 60.1 & 0.73\\
\enddata
\tablecomments{The notation of spurs as provided in column (1) is 
indicated in Fig.\,\ref{fig:region}. The number of objects found is listed in column (2). 
The remaining columns give the mean value of the GMC radius (3), the line width (4), 
the $\rm H_2$ gas mass (5) derived from the CO(1-0) luminosity and corrected for He 
contribution as listed in the GMC catalog of \citet{colombo14a}, 
the gas mass surface density per area analyzed (6), the total molecular gas mass in the
GMCs (7) and compared to the total molecular gas mass from Tab. \ref{tab:co} (8).
The properties of the individual GMCs are 
listed in Tab. \ref{tab:gmcs}. 
The mean for all properties of GMCs found in the individual spurs is given in the second 
last row, the corresponding properties for the arm segment in the last row. 
Values  are derived assuming a distance to M51a of 7.6\,Mpc and a Galactic 
conversion factor of $\rm 2x10^{20}\,cm^{-2}K^{-1}km^{-1}s$ for CO intensity into \htwo ~gas mass.  
\label{tab:gmcs_sum}}
\end{deluxetable}

\clearpage


\begin{deluxetable}{c|cccc}
\tabletypesize{\normalsize}
\tablecaption{Properties of HII Regions located in Northern Spiral Segment}
\tablehead{
\colhead{Spur} &
\multicolumn{4}{c}{HII region}
\\
\colhead{ID} &
\colhead{\#}&
\colhead{\it r} &
\colhead{$\rm log({\it L}_{H\alpha})$} &
\colhead{$\rm log({\it L}_{H\alpha})/{\it A}$} 
\\
& &
\colhead{(pc)} &
\colhead{($\rm log(erg\,s^{-1}$))} &
\colhead{($\rm log(erg\,s^{-1}/pc^2)$)} 
}
\startdata
1 &   3 & 35 & 37.31 & 33.11 \\     
2 &   4 & 50 & 38.00 & 33.80 \\    
3 &   0 & -- & ---   &  --- \\
4 &   2 &102 & 38.89 & 34.55 \\    
5 &   2 & 35 & 37.33 & 33.01 \\ 
6 &   1 &188 & 39.47 & 34.54 \\    
7 &   3 & 53 & 38.02 & 33.85 \\    
8 &   3 & 40 & 37.55 & 33.10 \\    
9 &   3 & 72 & 38.65 & 34.08 \\ \hline
mean & 2.3 & 72 & 38.75 & 34.09 \\ \hline
arm & 5 & 35 & 37.55 & 32.63 \\
\enddata
\tablecomments{The notation of spurs as provided in column (1) is 
indicated in Fig.\,\ref{fig:region}. The number of objects found is listed in column (2). 
The remaining columns give the mean value of the HII region radius in column (3), the logarithm of the
H$\alpha$ luminosity $\rm {\it L}_{H\alpha}$ in column (4), and the $\rm {\it L}_{H\alpha}$ per area analyzed in column (5). 
The properties of the individual HII regions are listed in Tab. \ref{tab:hii} and are taken from
the group catalog of \citet{lee11}. 
The mean for all properties of HII regions found in the individual spurs is given in the second 
last row, the corresponding properties for the arm segment in the last row. 
Values  are derived assuming a distance to M51a of 7.6\,Mpc and the properties of the HII regions
have been corrected correspondingly.  
\label{tab:hii_sum}}
\end{deluxetable}

\clearpage


\begin{deluxetable}{c|cccc}
\tabletypesize{\normalsize}
\tablecaption{Properties of Stellar Clusters located in Northern Spiral Segment}
\tablehead{
\colhead{Spur} &
\multicolumn{4}{c}{stellar cluster}
\\
\colhead{ID} &
\colhead{\#} &
\colhead{$\rm log({\it t})$} &
\colhead{$\rm {\it M}_{\star}$} &
\colhead{$\rm {\it M}_{\star}/{\it A}$} 
\\
& &
\colhead{($\rm log(yr)$)} &
\colhead{($\rm 10^4\,M_{\odot}$)} &
\colhead{($\rm M_{\odot}/pc^2$)} 
}
\startdata
1 &  3 & 6.74 & 0.5 & 0.3  \\     
2 &   5 & 6.60 & 1.5  & 1.2 \\    
3 &   0 & ---  & --- & ---  \\
4 &   5 & 6.54 & 1.8 & 2.1  \\    
5 &   1 & 6.56 & 0.4 & 0.1   \\ 
6 & 10 & 6.42 & 2.0 & 2.4   \\    
7 &   1 & 6.00 & 3.4 & 0.8  \\    
8 &   1 & 6.78 & 0.1 & 0.02  \\    
9 &   2 & 6.01 & 2.7 & 0.5   \\ \hline
mean & 3.1 & 6.46 & 1.5 & 0.9  \\ \hline
arm &  6 & 6.64 & 0.7 & 0.02  \\
\enddata
\tablecomments{The notation of spurs as provided in column (1) is 
indicated in Fig.\,\ref{fig:region}. The number of young ($\rm \le 10\,Myr$ stellar clusters found is listed in column (2). 
The remaining columns give the logarithm of the mean age in column (3), the mean stellar mass in
column (4), and the mean stellar mass density (from clusters) in column (5).
The properties of the individual stellar clusters are listed in Tab. \ref{tab:clusters} and are taken from
the catalog of \citet{chandar16}. 
The mean for all properties of young stellar clusters found in the individual spurs is given in the second last column, the
corresponding properties for the arm segment in the last row. 
Values 
are derived assuming a distance to M51a of 7.6\,Mpc and the properties of the stellar clusters
have been corrected correspondingly.  
\label{tab:cluster_sum}}
\end{deluxetable}

\clearpage


\begin{deluxetable}{c|cccccc}
\tabletypesize{\normalsize}
\tablecaption{GMCs identified in Spurs of the Northern Spiral Segment}
\tablehead{
\colhead{Spur} &
\colhead{ID} &
\colhead{RA(J2000)} &
\colhead{Dec.(J2000)} &
\colhead{\it r} &
\colhead{$\rm \Delta\,{\it v}$} &
\colhead{$\rm {\it M}_{H_2}$} 
\\
& &
\colhead{(deg)} &
\colhead{(deg)} &
\colhead{(pc)} &
\colhead{(km/s)}&
\colhead{($\rm 10^5\,M_{\odot}$)}
}
\startdata
  1 & 1291 & 202.4786149 & 47.2077502  & 33 &  4.7 &   9.86 \\
    & 1292 & 202.4803521 & 47.2092043  & 44 &  4.9 &   7.93 \\
    & 1295 & 202.4792608 & 47.2082393  & 39 & 16.0 &  11.85 \\
    & 1296 & 202.4793004 & 47.2088399  & 36 &  5.7 &  14.35 \\ \hline
  2 & 1285 & 202.4778061 & 47.2096689  & 47 &  1.2 &   6.89 \\
    & 1286 & 202.4779085 & 47.2102934  & 40 &  7.9 &  19.47 \\
    & 1298 & 202.4782993 & 47.2119982  & 26 &  7.8 &   3.26 \\ \hline
  3 & 1344 & 202.4754305 & 47.2108011  & 30 &  9.3 &   3.49 \\
    & 1354 & 202.4748221 & 47.2102946  & 61 &  5.8 &  21.17 \\
    & 1379 & 202.4749249 & 47.2110108  & 53 &  2.4 &   6.23 \\ \hline
  4 & 1356 & 202.4718143 & 47.2113143 1& 23 &  4.0 &  61.53 \\ \hline
  5 & 1358 & 202.4694169 & 47.2122155  & 60 &  4.4 &  29.90 \\ \hline
  6 & 1348 & 202.4677529 & 47.2119800  & 53 & 10.0 &  44.63 \\
    & 1349 & 202.4677036 & 47.2128936  & 75 &  6.8 &  49.56 \\
    & 1357 & 202.4668129 & 47.2118519  & 59 &  5.4 &  34.43 \\ \hline
  7 & 1406 & 202.4629751 & 47.2123380  & 38 &  5.7 &  10.05 \\
    & 1410 & 202.4624958 & 47.2117164  & 52 &  7.2 &  31.37 \\
    & 1419 & 202.4625185 & 47.2113040  & 74 &  1.8 &  35.43 \\ \hline
  8 & 1411 & 202.4588562 & 47.2133714 1& 25 &  8.2 &  62.20 \\ \hline
  9 & 1399 & 202.4543621 & 47.2109655  & 42 &  5.1 &  10.36 \\
    & 1401 & 202.4560851 & 47.2131609  & 16 &  5.0 &   5.45 \\
    & 1405 & 202.4558906 & 47.2114785  & 78 & 12.0 &  51.44 \\
    & 1412 & 202.4554843 & 47.2134147  & 53 &  1.7 &   7.57 \\
    & 1420 & 202.4549731 & 47.2124528  & 45 &  5.4 &   9.28 \\
    & 1421 & 202.4546728 & 47.2138573  & 38 &  4.4 &   2.91 \\ 
\enddata
\tablecomments{The notation of spurs as provided in column (1) is indicated in 
Fig.\,\ref{fig:region}. 
The remaining columns give the identification number (2), the position (3,4), the GMC radius (5), the line width (6), 
and the $\rm H_2$ gas mass (7) derived from the CO(1-0) luminosity and corrected for He 
contribution as listed in the GMC catalog of \citet{colombo14a}.  
\label{tab:gmcs}}
\end{deluxetable}


\begin{deluxetable}{cccccc}
\tabletypesize{\normalsize}
\tablecaption{GMCs identified in Arm Segment of the Northern Spiral}
\tablehead{
\colhead{ID} &
\colhead{RA(J2000)} &
\colhead{Dec.(J2000)} &
\colhead{\it r} &
\colhead{$\rm \Delta\,{\it v}$} &
\colhead{$\rm {\it M}_{H_2}$} 
\\
&
\colhead{(deg)} &
\colhead{(deg)} &
\colhead{(pc)} &
\colhead{(km/s)}&
\colhead{($\rm 10^5\,M_{\odot}$)}
}
\startdata
 1288 & 202.4781341 & 47.2064604  & 79 &  7.9 &  89.21 \\
 1290 & 202.4775177 & 47.2075980  & 24 & 18.6 &  33.84 \\
 1294 & 202.4794342 & 47.2058954  & 77 &  8.4 &  26.36 \\
 1301 & 202.4783715 & 47.2070463  & 38 &  5.1 &   7.82 \\
 1307 & 202.4778494 & 47.2064959  & 36 &  6.6 &   5.48 \\
 1340 & 202.4724277 & 47.2092676  &  0 &  4.2 &   0.98 \\
 1346 & 202.4720983 & 47.2101305  & 59 & 11.5 &  32.37 \\
 1347 & 202.4657128 & 47.2105135  & 35 &  4.9 &  12.34 \\
 1352 & 202.4754194 & 47.2094228  & 72 &  6.9 &  29.01 \\
 1353 & 202.4737470 & 47.2096134  & 73 &  5.8 &  29.44 \\
 1355 & 202.4705114 & 47.2108802  & 28 &  4.6 &   1.43 \\
 1365 & 202.4769357 & 47.2077326  & 62 &  6.1 &  16.35 \\
 1367 & 202.4750843 & 47.2089325  & 48 &  1.8 &   8.57 \\
 1368 & 202.4699346 & 47.2103279  & 62 & 12.5 &  49.36 \\
 1375 & 202.4761133 & 47.2084144  & 44 &  9.2 &  15.29 \\
 1377 & 202.4682569 & 47.2107920 1& 11 &  7.2 &  79.60 \\
 1378 & 202.4661434 & 47.2108190  & 25 & 11.3 &  35.56 \\
 1381 & 202.4764315 & 47.2069278  & 62 &  6.4 &  33.85 \\
 1388 & 202.4725072 & 47.2090741  & 42 &  6.7 &   5.24 \\
 1400 & 202.4633815 & 47.2105062  & 22 &  2.2 &   1.93 \\
 1409 & 202.4564362 & 47.2109857  & 48 &  6.7 &  19.54 \\
 1417 & 202.4554359 & 47.2094391  & 69 & 12.1 &  28.28 \\
 1426 & 202.4599843 & 47.2110239  & 39 &  5.1 &  15.02 \\
 1432 & 202.4561624 & 47.2101547  & 38 &  9.6 &  13.58 \\
 1433 & 202.4609935 & 47.2107303  & 64 &  7.5 &  30.85 \\
 1442 & 202.4549386 & 47.2098358  & 36 &  4.5 &   4.00 \\
 1448 & 202.4618436 & 47.2103964  & 39 &  1.8 &   5.30 \\
\enddata
\tablecomments{
The columns give the identification number (1), the position (2, 3), the GMC radius (4), the line width (5), 
and the $\rm H_2$ gas mass (6) derived from the CO(1-0) luminosity and corrected for He 
contribution as listed in the GMC catalog of \citet{colombo14a}.  
\label{tab:gmcs_arm}}
\end{deluxetable}


\begin{deluxetable}{c|ccccc}
\tabletypesize{\normalsize}
\tablecaption{HII regions identified in Northern Spiral Segment}
\tablehead{
\colhead{Region} &
\colhead{ID} &
\colhead{RA(J2000)} &
\colhead{Dec.(J2000)} &
\colhead{$\rm log({\it L}_{H\alpha})$} &
\colhead{\it r} 
\\
& &
\colhead{(deg)} &
\colhead{(deg)} &
\colhead{($\rm log(erg\,s^{-1})$)}&
\colhead{(pc)} 
}
\startdata
  1 & 10901 & 202.4788818 & 47.208076 &  37.475 &  44.22 \\ 
    & 11070 & 202.4794006 & 47.208359 &  37.195 &  35.74 \\ 
    & 11086 & 202.4794159 & 47.208946 &  37.198 &  23.58 \\ \hline
  2 & 10690 & 202.4779358 & 47.209621 &  37.142 &  30.58 \\ 
    & 10450 & 202.4778290 & 47.209972 &  38.519 & 104.64 \\ 
    & 10932 & 202.4788055 & 47.210018 &  37.218 &  22.48 \\ 
    & 10816 & 202.4784851 & 47.211365 &  37.585 &  41.64 \\ \hline
  4 &  8291 & 202.4715424 & 47.211002 &  39.181 & 162.49 \\
    &  8786 & 202.4718170 & 47.211468 &  37.431 &  42.37 \\ \hline
  5 &  7881 & 202.4695282 & 47.212265 &  37.218 &  30.21 \\ 
    &  8100 & 202.4700165 & 47.212067 &  37.427 &  39.79 \\ \hline
  6 &  6795 & 202.4669800 & 47.212589 &  39.471 & 188.28 \\ \hline
  7 &  5732 & 202.4627228 & 47.211338 &  37.695 &  40.16 \\ 
    &  5902 & 202.4635162 & 47.212296 &  37.572 &  47.16 \\ 
    &  5841 & 202.4631805 & 47.212852 &  38.354 &  71.11 \\ \hline
  8 &  4690 & 202.4594727 & 47.212654 &  37.637 &  46.43 \\ 
    &  4540 & 202.4587860 & 47.213089 &  37.413 &  32.79 \\ 
    &  4554 & 202.4589386 & 47.214386 &  37.566 &  39.43 \\ \hline
  9 &  4088 & 202.4557800 & 47.211254 &  39.101 & 129.70 \\
    &  4121 & 202.4561462 & 47.212753 &  37.850 &  55.27 \\ 
    &  4010 & 202.4547729 & 47.211903 &  37.349 &  30.95 \\ \hline
arm & 10634 & 202.4779205 & 47.206657 &  37.326 &  33.16 \\ 
    & 10755 & 202.4781647 & 47.207058 &  37.399 &  31.69 \\ 
    &  9797 & 202.4753571 & 47.209400 &  37.476 &  40.53 \\ 
    &  7346 & 202.4682922 & 47.210773 &  37.943 &  42.74 \\ 
    &  8213 & 202.4701996 & 47.210915 &  37.147 &  27.27 \\
\enddata    
\tablecomments{The notation of spurs as provided in column (1) is 
indicated in Fig.\,\ref{fig:region}. The remaining columns give the identification number (2),  
the position (3,4), the logarithm of the
H$\alpha$ luminosity $\rm {\it L}_{H\alpha}$ (5)
and the HII region radius (6) from
the HII region group catalog of \citet{lee11}. 
Values  are derived assuming a distance to M51a of 7.6\,Mpc and the properties of the HII regions
have been corrected correspondingly.  
\label{tab:hii}}
\end{deluxetable}


\begin{deluxetable}{c|ccccc}
\tabletypesize{\normalsize}
\tablecaption{Stellar Clusters identified in Northern Spiral Segment}
\tablehead{
\colhead{Region} &
\colhead{ID} &
\colhead{RA(J2000)} &
\colhead{Dec.(J2000)} &
\colhead{log({\it t})} &
\colhead{$\rm {\it M}_{\star}$} 
\\
& &
\colhead{(deg)} &
\colhead{(deg)} &
\colhead{($\rm log(yr)$)} &
\colhead{($\rm 10^3\,M_{\odot}$)} 
}
\startdata
  1 &   149411 & 202.4790344 & 47.2080193 & 6.56  &  7.0  \\ 
    &   150175 & 202.4795380 & 47.2084045 & 6.78  &  6.4  \\ 
    &   153018 & 202.4797974 & 47.2097244 & 6.88  &  1.6  \\ \hline
  2 &   155581 & 202.4789276 & 47.2108917 & 6.84  & 21.5  \\ 
    &   153350 & 202.4769440 & 47.2098770 & 6.84  & 39.0  \\ 
    &   152222 & 202.4772186 & 47.2093582 & 6.90  &  1.8  \\ 
    &   154051 & 202.4785461 & 47.2101517 & 6.44  &  4.7  \\ 
    &   154498 & 202.4782562 & 47.2103729 & 6.00  &  9.2  \\ \hline
  4 &   155911 & 202.4720154 & 47.2110329 & 6.38  & 18.7  \\ 
    &   155822 & 202.4718170 & 47.2109871 & 6.58  & 51.7  \\ 
    &   155199 & 202.4712219 & 47.2107048 & 6.20  & 18.5  \\ 
    &   157826 & 202.4706268 & 47.2119026 & 6.78  &  1.3  \\ 
    &   156475 & 202.4728851 & 47.2112885 & 6.78  &  1.3  \\ \hline
  5 &   158075 & 202.4695892 & 47.2120247 & 6.56  &  3.4  \\ \hline
  6 &   157786 & 202.4667053 & 47.2118950 & 6.00  & 52.3  \\ 
    &   158632 & 202.4673309 & 47.2122879 & 6.38  & 22.2  \\ 
    &   159154 & 202.4674835 & 47.2125168 & 6.46  & 26.3  \\ 
    &   159671 & 202.4675598 & 47.2127380 & 6.52  & 35.7  \\ 
    &   160624 & 202.4669189 & 47.2131500 & 6.02  & 46.5  \\ 
    &   159188 & 202.4670715 & 47.2125359 & 6.56  &  5.7  \\ 
    &   159729 & 202.4670563 & 47.2127609 & 6.54  &  4.4  \\ 
    &   161017 & 202.4665375 & 47.2133713 & 6.00  &  8.4  \\ 
    &   160512 & 202.4660034 & 47.2130966 & 6.98  &  2.5  \\ 
    &   160542 & 202.4685059 & 47.2131195 & 6.78  &  0.8  \\ \hline
  7 &   159521 & 202.4632874 & 47.2126770 & 6.00  & 34.4  \\ \hline
  8 &   159490 & 202.4597321 & 47.2126503 & 6.78  &  1.3  \\ \hline
  9 &   156907 & 202.4561462 & 47.2114792 & 6.00  & 21.0  \\ 
    &   156216 & 202.4559174 & 47.2111664 & 6.02  & 32.2  \\ \hline
arm &   152028 & 202.4754333 & 47.2092552 & 6.68  & 12.9  \\ 
    &   146559 & 202.4779510 & 47.2066536 & 6.02  &  7.2  \\ 
    &   146174 & 202.4772491 & 47.2064781 & 6.78  &  2.8  \\ 
    &   145563 & 202.4776611 & 47.2061653 & 6.76  & 13.3  \\
    &   152720 & 202.4737396 & 47.2095757 & 6.76  &  2.7  \\ 
    &   155880 & 202.4681549 & 47.2110176 & 6.48  &  3.3  \\ 
\enddata
\tablecomments{The notation of spurs as provided in column (1) is 
indicated in Fig.\,\ref{fig:region}. 
The remaining columns give the identification number (2), the position (3,4), the logarithm of the age (5), 
and the stellar mass (6) of the young ($\rm {\it t} \le 10\,Myr$) stellar clusters in the catalog of \citet{chandar16}. 
Values 
are derived assuming a distance to M51a of 7.6\,Mpc and the properties of the stellar clusters
have been corrected correspondingly.  
\label{tab:clusters}}
\end{deluxetable}
    

\clearpage


\clearpage


\clearpage


\clearpage


\clearpage


\clearpage


\begin{figure}[ht]
\begin{center}
\resizebox{.9\hsize}{!}{\includegraphics[angle=-90]{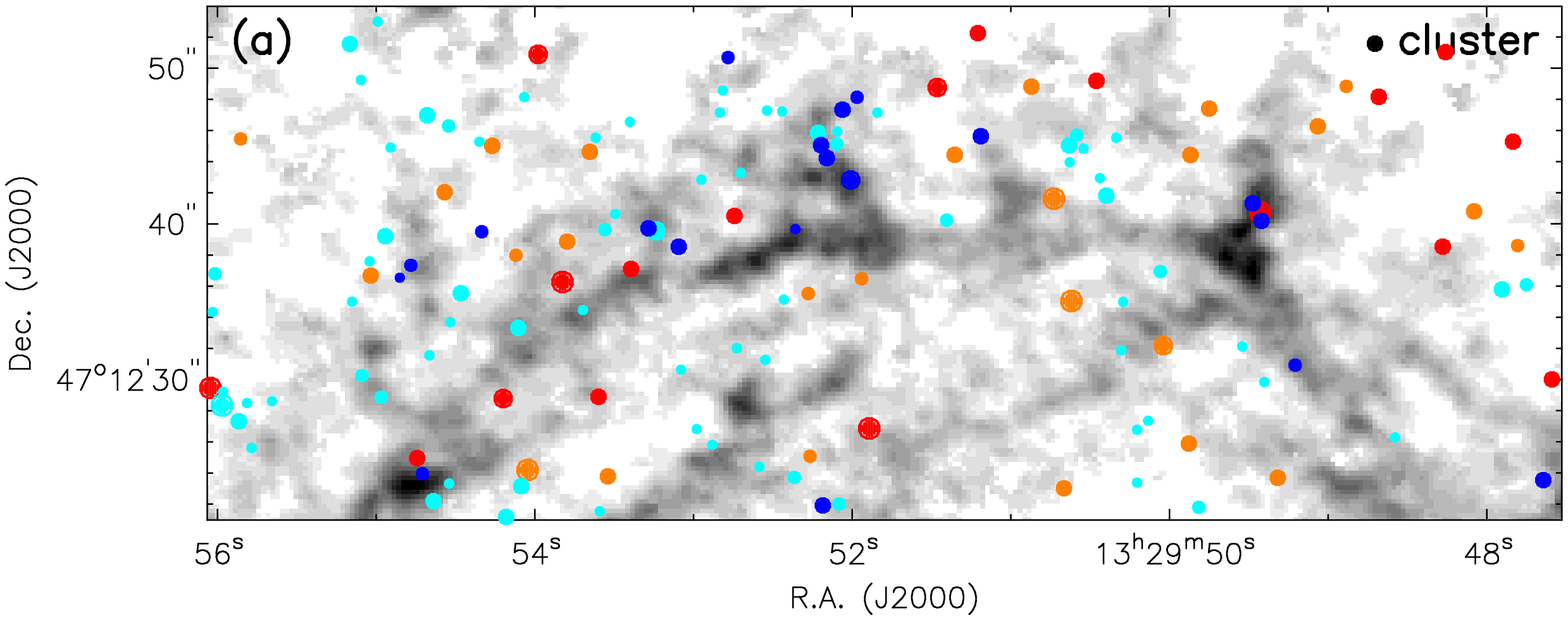}}
\end{center}
\caption{Detailed comparison of the location of stellar clusters (filled circles) relative to the molecular gas 
distribution (greyscale). The stellar clusters are shown by filled circles symbols (same size as for 
Fig.\,\ref{fig:spur_obj}c). The color coding corresponds to ages of 
$\rm log{\it  t}(yr)) < 6.5 ~(blue), 
6.5 \le log({\it t}(yr)) < 7.0 ~(cyan), 
7.0 \le log({\it t}(yr)) < 8.2 ~(orange), and ~
log({\it t}(yr)) \ge 8.2 ~(red)$.
It is noteworthy that clusters with ages of $\rm log({\it t}(yr)) < 6.5$ tend to be highly clusters, while the clustering is
becoming less obvious for clusters with ages of $\rm 6.5 \le log({\it t}(yr)) < 7.0$ and is non-apparent for even older
clusters.
\label{fig:clusters}}
\end{figure}


\clearpage


\end{document}